\documentclass[11pt]{article}

\usepackage{algorithm}
\usepackage{algorithmic}
\usepackage{epsfig,subfigure}
\usepackage{amssymb, amsmath}
\usepackage{color}
\usepackage{setspace}
\usepackage{graphicx}

\usepackage{amssymb}
\usepackage{amsmath,amsfonts,amssymb}
\usepackage{verbatim}
\usepackage[bookmarks=false]{}
\usepackage[font={small}]{caption}

\newtheorem{theorem}{Theorem}

\newtheorem{lemma}{Lemma}

\newtheorem{remark}{Remark}

\newtheorem{example}{Example}

\begin{document}
\title{Multilevel Topological Interference Management: A TIM-TIN Perspective}
\date{}
\author{Chunhua Geng, Hua Sun, Syed A. Jafar% <-this % stops a space
\thanks{Chunhua Geng (email: chunhua.geng@mediatek.com) is with MediaTek USA Inc., Irvine, CA. Hua Sun (email: hua.sun@unt.edu) is with the Department of Electrical Engineering, University of North Texas, Denton, TX.  Syed A. Jafar (email: syed@uci.edu) is with the Center of Pervasive Communications and Computing (CPCC) in the Department of Electrical Engineering and Computer Science (EECS) at the University of California Irvine, Irvine, CA. A part of this work was presented in IEEE Information Theory Workshop (ITW) 2013 \cite{Geng_MTIM}. }
}

\maketitle

\begin{abstract}
The robust principles of treating interference as noise (TIN) when it is sufficiently weak, and avoiding it when it is not, form the background of this work. Combining TIN with the topological interference management (TIM) framework that identifies optimal interference avoidance schemes, we formulate a TIM-TIN problem for multilevel topological interference management, wherein only a coarse knowledge of channel strengths and no knowledge of channel phases is available to transmitters. To address the TIM-TIN problem, we first propose an analytical baseline approach, which decomposes a network into TIN and TIM components, allocates the signal power levels  to each user in the TIN component, allocates signal vector space dimensions  to each user in the TIM component, and guarantees that the product of the two is an achievable number of signal dimensions available to each user in the original network. Next, a distributed numerical algorithm called ZEST is developed. The convergence of the algorithm is demonstrated, leading to the duality of the TIM-TIN problem (in terms of GDoF). Numerical results are also provided to demonstrate the superior sum-rate performance and fast convergence of ZEST. \end{abstract}

%\begin{IEEEkeywords}
%Capacity region, Gaussian interference channel, generalized degrees of freedom (GDoF), treating interference as noise (TIN).
%\end{IEEEkeywords}

% \fi

\section{Introduction}
The capacity of wireless interference networks is a rapidly evolving research front, spurred in part by exciting breakthroughs such as the idea of interference alignment \cite{Jafar_FnT} which provides fascinating theoretical insights and shows much promise  under idealized conditions. The connection to practical settings however remains tenuous. This is in part due to the following two factors. First,  because of the assumption of precise channel knowledge, idealized studies often get caught in the minutiae of channel realizations, e.g., rational versus irrational values, that have little bearing in practice. Second, by focusing on the degrees-of-freedom (DoF) of fully connected networks,  these studies ignore the most critical aspect of interference management in practice -- the differences of signal strengths due to path loss and fading (in short, network topology). Indeed, the DoF metric treats every channel as essentially equally strong (capable of carrying exactly 1 DoF). So the desired signal has to actively avoid \emph{every} interferer, whereas in practice each user needs to avoid only  a few significant interferers and the rest are weak enough to be safely ignored. Therefore, by trivializing the topology of the network, the DoF studies of fully connected networks make the problem much harder than it needs to be. Non-trivial solutions to this harder problem  invariably rely on much more channel knowledge than is available in practice. Thus, the two limiting factors re-enforce each other.

Evidently, in order to avoid these pitfalls, one should shift focus away from optimal ways of exploiting precise channel knowledge (which is rarely available), and toward powerful even optimal ways of exploiting a coarse knowledge of interference network \emph{topology}. This line of thought motivates robust models of interference networks  where  only a {coarse} knowledge of channel strength {levels} is available to the transmitters and no channel phase knowledge is assumed. This is the \emph{multilevel} topological interference management framework. It is a generalization of the elementary topological interference management (TIM) framework introduced in \cite{Jafar_TIM}, wherein the transmitters can only distinguish between channels that are connected (strong) and not connected (weak).
\vspace{-0.18in}

\subsection{Robust principles of interference management: Ignore, avoid}
Existing wireless interference networks are mainly based on two robust interference management principles ---  1) \emph{ignore}  interference that is sufficiently weak, and 2) \emph{avoid}  interference that is not. In slightly more technical terms, ignoring interference translates into \emph{treating interference as noise} (TIN) \cite{Chara_TIN,BinaryPC}, and avoiding interference translates into access schemes such as TDMA/FDMA/CDMA. Recent work has explored the optimality of both of these principles.

\begin{enumerate}
\item {\it TIN:} The optimality of the first principle, treating interference as noise when it is sufficiently weak, is discussed extensively. In \cite{Annapureddy_Veeravalli_TIN_opt,Motahari_Khandani_TIN_opt,Kramer_TIN_opt}, it is shown that in a so-called ``noisy interference'' regime, TIN achieves the exact \emph{sum} capacity of interference channels. In \cite{Geng_TIN}, for general $K$-user interference channels, it provides a broadly applicable TIN-optimality condition under which TIN is optimal from a generalized degrees-of-freedom (GDoF) perspective and achieves a constant gap (of no more than $\log(3K)$ bits) to the entire capacity region. More specifically, under a fully asymmetric setting, the TIN-optimality condition identified in \cite{Geng_TIN} stipulates that if for each user the desired signal strength is no less than the sum of the strongest interference \emph{from} this user and the strongest interference \emph{to} this user (all values in dB scale), then power control and TIN achieves the whole GDoF region of this network. Remarkably, this result holds even if perfect channel knowledge is assumed everywhere. The TIN-optimality result is also generalized to other channel models (e.g., $X$ channels \cite{Geng_TIN_X, Aydin_TIN_X}, parallel channels \cite{Sun_TIN_Parallel}, compound networks \cite{Geng_PC}, MIMO channels \cite{Geng_MIMOTIN}, and cellular networks \cite{Cellular_uplink, Cellular_downlink}) and is reformulated from a combinatorial perspective \cite{Yi_PC}.

% which subsumes all the previous results in the symmetric settings as special cases \cite{Tse_GDoF,Jafar_Vishwanath_GDoF}. 
\item {\it TIM:} The optimality of the second principle, \emph{avoidance}, has been investigated most recently by \cite{Jafar_TIM}, as the TIM problem. With channel knowledge at the transmitters limited to a coarse knowledge of network topology (which links are stronger/weaker than the effective noise floor),  TIM  is shown in \cite{Jafar_TIM} to be essentially  an index coding problem \cite{Birk_Kol}.  TIM subsumes within itself the TDMA/FDMA/CDMA schemes as trivial special cases, but is in general much more capable  than these conventional approaches. Remarkably, for the class of linear schemes, which are found to be optimal in most cases studied so far, and within which TIM is equivalent to the index coding problem, TIM is essentially an optimal allocation of signal vector spaces based on an interference alignment perspective \cite{Maleki_Cadambe_Jafar_index}. Variants of the TIM problem have also been investigated, such as those under short coherent time \cite{Navid_TIM}, with alternating connectivity  \cite{Sun_ATIM, Aydin_ATIM}, with multiple antennas \cite{Sun_MIMOTIM}, with transmitter/receiver cooperation \cite{Yi_coTIM, Yi_deTIM}, with reconfigurable antennas \cite{TIM_reAnt}, with network topology uncertainty \cite{Oppo_TIM}, and with confidential messages \cite{secure_TIM}.
\end{enumerate}
\vspace{-0.16in}

\subsection{TIM-TIN: Joint view of signal vector spaces and signal power levels}
The two principles -- avoiding versus ignoring interference -- which are mapped to TIM and TIN, respectively, naturally correspond to interference management in terms of signal  \emph{vector spaces} and signal \emph{power levels}. TIM uses the interference alignment perspective \cite{Jafar_TIM, Maleki_Cadambe_Jafar_index} to optimally allocate signal vector subspaces among the interferers. Note that in order to resolve the desired signal from interference based on the signal vector spaces, the strength of each signal is irrelevant. What matters is only that desired signal and the interference occupy linearly independent spaces. TIN, on the other hand, optimally allocates signal power levels among users by setting the transmit power levels at transmitters and the noise floor levels at receivers. Thus TIN depends very much on the strengths of signals relative to each other. Associating TIM with signal vector space allocations and TIN with signal power level allocations within the multilevel TIM framework, we refer to the joint allocation of signal vector spaces and signal power levels as the TIM-TIN problem.

%While  interference management approaches  based on signal spaces and signal levels have each been extensively studied in a variety of settings, combining the two has been a challenge (a notable recent work is \cite{Cadambe_Nazer_Caire}), especially when the schemes involved are rather fragile because of their extreme sensitivity to the precise channel realizations. However, because of the minimal channel knowledge assumptions in the TIM and TIN settings,  perhaps relatively simpler and robust (even if sub-optimal) algorithms may naturally arise for the joint allocations of signal spaces and signal levels. Within the multilevel topological interference management framework, this joint TIM-TIN problem is the focus of this work.

{\it TIM-TIN Problem:} With only a coarse knowledge of channel strengths available to the transmitters, we wish to carefully allocate not only the beamforming vector {directions} (signal vector spaces) but also the transmit powers (signal power levels) to each of those beamforming vectors. The necessity of a joint TIM-TIN perspective is evident as follows. In vector space allocation schemes used for DoF studies, the signal space containing the interference is entirely rejected (zero-forced). This is typically fine for linear DoF studies because all signals are essentially equally strong,  every substream carries one DoF, so any desired signal projected into the interference space cannot achieve a non-zero DoF. However, once we account for the difference in signal strengths in the GDoF framework,  the signal vector space dimensions occupied by interference may not be \emph{fully} occupied in terms of power levels if the interference is weak. So, non-zero GDoF may be achieved by desired signals projected into the same dimensions as occupied by the interference, where interference is weaker than desired signal. It is this aspect that we wish to exploit in this work. It is worthwhile noticing that within the multilevel TIM framework, in general the solution based on a combination of TIM and TIN is not optimal. In \cite{Arash_GDoF},  it has been shown that  for $K$-user \emph{symmetric} interference channels, the GDoF optimal solution relies on rate splitting and superposition encoding at transmitters and (partial) interference decoding at receivers.\footnote{Within the multilevel TIM framework, for a $K$-user interference with arbitrary channel strengths, the optimal GDoF region is still open.} The appeal of joint TIM-TIN mainly lies in its implementation simplicity and wide applicability in existing wireless networks.

\subsection{Overview of results}
First, to address the TIM-TIN problem, an analytical baseline approach is presented. Because of the minimal channel knowledge requirements in the TIM and TIN settings, a robust combination of the two, denoted as \emph{TIM-TIN decomposition} presents itself. Any given network is decomposed into a TIM component and a TIN component, containing only  strong and weak interferers, respectively, and a direct multiplication of the signal dimensions available in each is shown to be achievable in the original network. In other words, the TIM solution identifies the fraction of the signal space that is available to each user, and within each of these available signal space dimensions, the TIN approach identifies the fraction of signal levels that are available to the same user. A product of the two fractions therefore identifies the net fraction of signal dimensions available to each user in this decomposition based approach. The optimality of this decomposition approach is also discussed for some non-trivial network settings. 

Next, a distributed numerical approach is developed for the TIM-TIN problem, which only needs local channel measurements to update transmit powers and beamforming vectors. The proposed algorithm, called ZEST, utilizes the reciprocity of wireless networks, and is guaranteed to be convergent in terms of GDoF. As a byproduct, the duality of the TIM-TIN problem is established. We also numerically validate the superior GDoF performance and fast convergence of ZEST.

\emph{Notations}: For a positive integer $Z$, $[Z]\triangleq\{1,2,...,Z\}$. $\mathbb{Z}^+$ and $\mathbb{Z}^-$ denote the sets of non-negative integers and non-positive integers, respectively. For vectors $\mathbf{u}$ and $\mathbf{v}$, we say that $\mathbf{u}$ dominates $\mathbf{v}$ if $\mathbf{u}\geq \mathbf{v}$, where $\geq$ denotes componentwise inequality. For a matrix $\mathbf{A}$, $\det(\mathbf{A})$ denotes its determinant, $\mathrm{span}(\mathbf{A})$ represents the space spanned by the column vectors of $\mathbf{A}$, and $\mathbf{A}(i,j)$ denotes the entry of $\mathbf{A}$ in the $i$-th row and $j$-th column. All logarithms are to the base $2$. %unless otherwise stated,

\section{System Model}\label{sec_model}
In this work, we consider a $K$-user complex Gaussian interference channel, where Transmitter $k$ ($k\in[K]$) intends to communicate with Receiver $k$  and all the transmitters and receivers are equipped with one antenna. Following \cite{Geng_TIN,Tse_GDoF}, the channel model is given by
\begin{align}
\label{equ_channel}
Y_k(t)=\sum_{i=1}^K\sqrt{P^{\alpha_{ki}}}e^{j\theta_{ki}}X_i(t)+Z_k(t),~~~ \forall k\in[K],
\end{align}
where at each time index $t$, $X_i(t)$ is the transmitted symbol of Transmitter $i$ (subject to a unit power constraint, i.e., $E[|X_i(t)|^2]\leq 1$), $Y_k(t)$ is the received signal of Receiver $k$, and $Z_k(t)\sim \mathcal{CN}(0,1)$ is the additive white Gaussian noise (AWGN) at Receiver $k$. In (\ref{equ_channel}), $P>1$ is a nominal power value, $\alpha_{ki}\geq 0$ is called the channel strength level of the link between Transmitter $i$ and Receiver $k$, and $\theta_{ki}$ is the corresponding channel phase. The definitions of messages, achievable rate of user $k$ ($R_k$) and channel capacity region ($\mathcal{C}$) are all standard.  The GDoF region is defined as
\begin{align}
\mathcal{D}\triangleq \Big\{(d_1,d_2,...,d_K): ~d_i=\lim_{P\rightarrow\infty}\frac{R_i}{\log P},~\forall i\in[K], ~(R_1,R_2,...,R_K)\in \mathcal{C}\Big\}.
\end{align}

In the multilevel TIM framework, only a {coarse} knowledge of channel strength {levels} is available to the transmitters and no channel phase knowledge is assumed. The channel strength level knowledge at transmitters can either be perfect or quantized. We also assume that receivers have perfect channel state information. Apparently, multilevel TIM is a generalization of the elementary one. It also should be noted that unlike most previous works in pursuit of the coarse DoF metric where  all non-zero channels are essentially treated as approximately equally strong (i.e., each non-zero channel carries one DoF), in the multilevel TIM framework, the main challenge lies in how to leverage the disparate channel strengths, and the more general GDoF metric is of interest. This progressive refinement (from DoF to GDoF) has been shown instrumental for capacity approximation of Gaussian interference networks in recent works \cite{Geng_TIN,Geng_TIN_X, Tse_GDoF,Karmakar_Varanasi_gap}, where the GDoF result usually further serves as a stepping stone for the capacity characterization within a constant gap.

Below we define the problem of multilevel TIM with \emph{quantized} channel strength levels, or quantized multilevel TIM (QM-TIM) in more details.\footnote{With a little abuse of notations, in QM-TIM, we also use $\alpha_{ij}$ to denote the quantized channel strength level for the link between Transmitter $j$ and Receiver $i$, $\forall i,j\in[K]$.}  Note that in practice, the desired signal strength and interfering signal strength usually fall into different ranges, so it is reasonable to assume that desired links and interfering links use different quantization levels. For direct channels, the channel strength levels are assumed to be large enough to guarantee a satisfying interference-free achievable rate.  As a result, for direct links the quantized channel strength levels are always normalized to one, i.e., $\alpha_{ii}=1$, $\forall i\in[K]$.\footnote{Here the unit quantized channel strength level is a result of normalization and imposes no loss of generality. More specifically, assume that we have multiple quantization levels for the direct links, and 1 is the highest quantization threshold, which represents a satisfying SNR value for the direct links (i.e., without interference, a satisfying achievable rate can be guaranteed). Through appropriate system design, it is natural to expect that all the direct links achieve this satisfying SNR value, thus the actual channel strength levels for all the direct links should be no less than 1.  After quantization we have $\alpha_{ii}=1$, $\forall i\in[K]$.} While for \emph{interfering links}, for better interference management, they are quantized by $l+1$ levels with quantization thresholds $t_1, t_2, ..., t_l$, where $0\leq t_1<t_2<...<t_l<\infty$. Hereafter, we denote the QM-TIM problem with the above quantization configuration by QM-TIM$(t_1,t_2,...,t_l)$. With those notations, the original TIM problem is a special case of QM-TIM, which can be denoted by QM-TIM(0). As another example, the simplest setting of  QM-TIM beyond the elementary one is QM-TIM$(t_1,t_2)$. One natural choice for the two quantized thresholds could be $t_1=0$ and $t_2=0.5$. In this case, we have the following three kinds of interfering links: 1) Weak interfering links: the interfering links that are no stronger than the noise floor; 2) Medium interfering links: the interfering links whose channel strength level value falls into the range from 0 to 0.5; 3) Strong interfering links: the interfering links whose channel strength level is no less than 0.5. %The set of all strong interfering links is denoted as $\mathcal{S}$.

\section{TIM-TIN Problem Formulation}\label{sec_timtin_form}
In this section, we formulate the TIM-TIN problem within the multilevel TIM framework formally. As mentioned before, with only a coarse knowledge of channel strengths available to the transmitters, in the TIM-TIN problem, we allocate not only the beamforming vectors but also the transmit powers to each of those beamforming vectors, in order to jointly optimize both signal vector space and signal power level allocations. 

For a $K$-user interference channel in (\ref{equ_channel}), over $n$ channel uses, Transmitter $i$ sends out $b_i$ ($b_i\leq n$) independent scalar data streams, each of which carries one symbol $s_{i,l}$ and is transmitted along an $n\times 1$ beamforming vector $\mathbf{v}_{i,l}$, $l\in[b_i]$. Assume that all symbols $s_{i,l}$ are drawn from independent Gaussian codebooks, each with zero mean and unit power, and the beamforming vectors $\mathbf{v}_{i,l}$ are scaled to have unit norm. Receiver $k$ obtains an $n\times 1$ vector over $n$ channel uses
\begin{align}
\mathbf{y}_k=\sum_{i=1}^K\sum_{l=1}^{b_i}\sqrt{P^{\alpha_{ki}}}e^{j\theta_{ki}}\sqrt{P^{r_{i,l}}}\mathbf{v}_{i,l}s_{i,l}+\mathbf{z}_k,
\end{align}
where  $\mathbf{z}_k$ is an $n\times 1$ zero mean unit variance circularly symmetric AWGN vector at Receiver $k$, $P^{r_{i,l}}$ is the transmit power for $l$-th data stream of User $i$. Due to the unit power constraint, we require  $r_{i,l}\leq 0$. For User $k$, the covariance matrix of the desired signal is
%\begin{align}
$\mathbf{Q}^D_k=\sum_{l=1}^{b_k}(\mathbf{v}_{k,l}\mathbf{v}_{k,l}^\dag)P^{r_{k,l}+\alpha_{kk}}$.
%\end{align}
The covariance matrix of the net interference-plus-noise is
%\begin{align}
$\mathbf{Q}^{N+I}_k=\sum_{i\neq k}\mathbf{Q}_{ki}+\mathbf{I},$
%\end{align}
where $\mathbf{I}$ is an $n\times n$ identity matrix, and $\mathbf{Q}_{ki}=\sum_{l=1}^{b_i}(\mathbf{v}_{i,l}\mathbf{v}_{i,l}^\dag)P^{r_{i,l}+\alpha_{ki}}$ is the covariance matrix of the interference from Transmitter $i\neq k$.
%\begin{align}
%$\mathbf{Q}_{ki}=\sum_{l=1}^{b_i}(\mathbf{v}_{i,l}\mathbf{v}_{i,l}^\dag)P^{r_{i,l}+\alpha_{ki}}$
%\end{align}
Given the beamforming vectors of each transmitter and power allocations of all data streams, \emph{as in the TIM-TIN problem the receivers do not attempt to decode interference from unintended transmitters}, for User $k\in[K]$ the achievable rate per channel use is given by
\begin{align*}
R_k&=\frac{1}{n}I(s_{k,1},s_{k,2},...,s_{k,b_k};\mathbf{y}_k)\\
&=\frac{1}{n}\Big[h(\mathbf{y}_k)-h(\mathbf{y}_k|s_{k,1},s_{k,2},...,s_{k,b_k})\Big]\\
&=\frac{1}{n}\bigg\{\log\Big[\det(\mathbf{Q}^D_k+\mathbf{Q}^{N+I}_k)\Big]-\log\Big[\det(\mathbf{Q}^{N+I}_k)\Big]\bigg\},
\end{align*}
and the achievable GDoF value $d_k$ is
\begin{align}
d_k=\lim_{P\rightarrow \infty}\frac{R_k}{\log P}=\lim_{P\rightarrow \infty}\frac{\log\Big[\det(\mathbf{Q}^D_k+\mathbf{Q}^{N+I}_k)\Big]-\log\Big[\det(\mathbf{Q}^{N+I}_k)\Big]}{n\log P}.\label{GDoF_general}
\end{align}

Next, we simplify the achievable GDoF expression into a more intuitive form.  Consider a term of the type $\log\Big[\det(\mathbf{I}+\sum_{i=1}^mP^{\kappa_i}\mathbf{v}_i\mathbf{v}_i^\dag)\Big]$, %$\log[\det(\mathbf{I}+P^{\kappa_1}\mathbf{v}_1\mathbf{v}_1^\dag+P^{\kappa_2}\mathbf{v}_2\mathbf{v}_2^\dag+...+P^{\kappa_m}\mathbf{v}_m\mathbf{v}_m^\dag)]$,
where $\mathbf{v}_i$ ($i\in[m]$) is an $n\times 1$ vector. Without loss of generality, assume $\kappa_1\geq\kappa_2\geq...\geq\kappa_m\geq0$. Consider the vectors $\mathbf{v}_i$'s one by one. For $\mathbf{v}_1$, we relabel it as $\mathbf{v}_{\Pi(1)}$ and correspondingly its power exponent $\kappa_{1}$ as $\kappa_{\Pi(1)}$. For $\mathbf{v}_2$, if it falls into $\mathrm{span}(\mathbf{v}_{\Pi(1)})$, we remove it and then proceed to $\mathbf{v}_{3}$; otherwise, we relabel it as $\mathbf{v}_{\Pi(2)}$ and correspondingly $\kappa_{2}$ as $\kappa_{\Pi(2)}$. We repeat this operation for each vector. In other words, for $\mathbf{v}_{i}$, if it falls into $\mathrm{span}(\mathbf{v}_{\Pi(1)}, \mathbf{v}_{\Pi(2)},...,\mathbf{v}_{\Pi(l)})$ (i.e., the space spanned by all previous linearly independent vectors obtained from $\{\mathbf{v}_1, \mathbf{v}_2,...,\mathbf{v}_{i-1}\}$), we remove it and then proceed to $\mathbf{v}_{i+1}$; otherwise, we relabel it as $\mathbf{v}_{\Pi(l+1)}$ and correspondingly its power exponent $\kappa_{i}$ as $\kappa_{\Pi(l+1)}$. Finally, we have $\gamma\leq n$ linearly independent beamforming vectors $\mathcal{V}_{\Pi} =\{\mathbf{v}_{\Pi(1)}, \mathbf{v}_{\Pi(2)},...,\mathbf{v}_{\Pi(\gamma)}\}$ and their associated power exponents $\mathcal{P}_{\Pi}=\{ \kappa_{\Pi(1)},\kappa_{\Pi(2)},...,\kappa_{\Pi(\gamma)}\}$.  With those definitions, we have the following lemma.

\begin{lemma}\label{Matrix_lemma}
Suppose that $\mathbf{v}_i$, $i\in[m]$ are $n\times 1$ vectors, and $\kappa_1\geq\kappa_2\geq...\geq\kappa_m\geq0$. We have
\begin{align}
\log\Big[\det(\mathbf{I}+\sum_{i=1}^mP^{\kappa_i}\mathbf{v}_i\mathbf{v}_i^\dag)\Big]
%&\log[\det(\mathbf{I}+P^{\kappa_1}\mathbf{v}_1\mathbf{v}_1^\dag+P^{\kappa_2}\mathbf{v}_2\mathbf{v}_2^\dag+...+P^{\kappa_m}\mathbf{v}_m\mathbf{v}_m^\dag)]
=\sum_{i=1}^\gamma \kappa_{\Pi(i)}\log P +o(\log(P)).%\nonumber
\end{align}
\end{lemma}
The proof of Lemma \ref{Matrix_lemma} is given in Appendix A. Now we can proceed to the following lemma.

\begin{lemma} \label{lemma_ZF}
In the TIM-TIN problem, given the beamforming vectors and the power allocations for each user, zero-forcing with successive cancellation (ZF-SC) achieves the maximal GDoF value of each user given by (\ref{GDoF_general}).\footnote{Note that with the ZF-SC receiver, each user only successively decodes and cancels the (possible multiple) desired data streams from its own transmitter, but does not decode interfering signals from others.}
\end{lemma}
The proof of Lemma \ref{lemma_ZF} is deferred to Appendix B. With Lemma \ref{lemma_ZF}, to maximize the achievable GDoF in the TIM-TIM problem,  the remaining challenge is choosing beamforming vectors and their powers for each user judiciously. To address this problem, in the following we develop two approaches, i.e., an analytical decomposition approach and a numerical distributed approach. %In the sequel of this paper, for given beamforming vectors and power allocations, we refer to the ZF-SIC receiver achieving the maximal GDoF as the 

\begin{figure}[h]
\begin{center}
 \includegraphics[width= 4 cm]{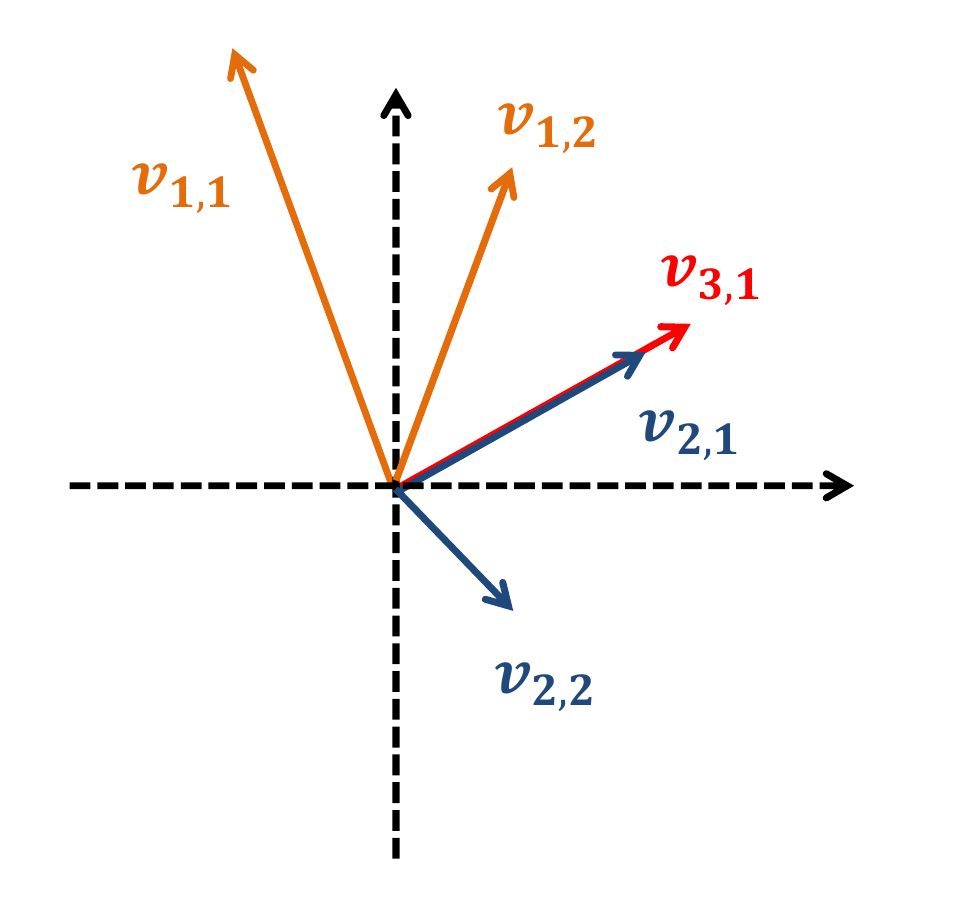}
 \caption{The received signal at Receiver 1, where the length of the vector represents the received power of the carried symbol. Here the number of channel uses $n$ is  2.}
\label{exp_GDoF}
\end{center}
\vspace{-0.3in}
\end{figure}

\begin{example}
To help understand Lemma \ref{Matrix_lemma} and \ref{lemma_ZF}, consider a $3$-user interference channel, in which over $2$ channel uses User $1$, $2$ and $3$ deliver $2$, $2$ and $1$ data streams, respectively.  Given the beamforming vectors, the transmitted power allocated to each symbol and channel strength levels for each link, the received signal at Receiver $1$ is depicted in Fig.~\ref{exp_GDoF}, where $\mathbf{v}_{2,1}$ and $\mathbf{v}_{3,1}$ are aligned along one direction. The length of the vector represents the received power of the carried symbol.  We have $r_{1,1}+\alpha_{11}>r_{1,2}+\alpha_{11}>r_{3,1}+\alpha_{13}>r_{2,1}+\alpha_{12}>r_{2,2}+\alpha_{12}>0.$ Define
$d_k'=\lim_{P\rightarrow \infty}\frac{\log[\det(\mathbf{Q}^D_k+\mathbf{Q}^{N+I}_k)]}{\log P}$ and 
$d_k''=\lim_{P\rightarrow \infty}\frac{\log[\det(\mathbf{Q}^{N+I}_k)]}{\log P}$. Following Lemma \ref{Matrix_lemma}, we have
$d_1'= r_{1,1}+\alpha_{11}+r_{1,2}+\alpha_{11}$ and 
$d_1''=r_{3,1}+\alpha_{13}+r_{2,2}+\alpha_{12}$.
So the achievable GDoF value of User $1$ is
\begin{equation}
\label{Exp_GDoF_equ}
\begin{aligned}
d_1=\frac{d_1'-d_1''}{2}=\frac{[(r_{1,1}+\alpha_{11}+r_{1,2}+\alpha_{11})-(r_{3,1}+\alpha_{13}+r_{2,2}+\alpha_{12})]}{2}
\end{aligned}
\end{equation}

Next, we illustrate how to achieve this GDoF value via a ZF-SC receiver. To decode $s_{1,1}$, we first zero force the strongest interference $s_{1,2}$ and then treat all the other interference as noise.  The achievable GDoF value of data stream $s_{1,1}$ is
\begin{align*}
d_{1,1}=\frac{(r_{1,1}+\alpha_{11}-\max\{r_{3,1}+\alpha_{13},r_{2,1}+\alpha_{12},r_{2,2}+\alpha_{12}\})}{2}=\frac{1}{2}(r_{1,1}+\alpha_{11}-r_{3,1}-\alpha_{13})
\end{align*}
After recovering $s_{1,1}$, we subtract it off from the received signal and then decode $s_{1,2}$. Similarly, we first zero force the strongest interference $s_{3,1}$ (and its aligned counterpart $s_{2,1}$) and then treat the remaining interference $s_{2,2}$ as noise. The achievable GDoF value of data stream $s_{1,2}$ is $d_{1,2}=\frac{1}{2}(r_{1,2}+\alpha_{11}-r_{2,2}-\alpha_{12})$. The achievable GDoF value for User $1$ is the sum of $d_{1,1}$ and $d_{1,2}$, which equals  (\ref{Exp_GDoF_equ}). Also note that the achievable GDoF value does not depend on the decoding order, i.e., if we reverse the decoding order of $s_{1,1}$ and $s_{1,2}$, we still achieve the same GDoF value for User 1.
\end{example}

\section{An Analytical Decomposition Approach}\label{sec_dcp}
%As shown in the previous subsection, the joint TIM-TIN problem is a challenging optimization problem over both signal vector space and signal levels. Due to the minimal channel knowledge requirements in both TIM and TIN settings, of these two approaches is promising. 

In this section, for the TIM-TIN problem we present an analytical baseline approach, denoted by \emph{TIM-TIN decomposition}.  The basic idea of this approach is as follows. Any given network can be decomposed into a TIM component and a TIN component, each containing all the desired links and non-overlapping interfering links, such that in total, these two components cover all the interfering links. In other words, denote the sets of all interfering links in the original network, TIM component, and TIN component by $\mathcal{I}$, $\mathcal{I}_{\mathrm{TIM}}$, and $\mathcal{I}_{\mathrm{TIN}}$, respectively.  We have $\mathcal{I}_{\mathrm{TIM}}\cap\mathcal{I}_{\mathrm{TIN}}=\phi$ and $\mathcal{I}_{\mathrm{TIM}}\cup\mathcal{I}_{\mathrm{TIN}}=\mathcal{I}$. 
First, consider the TIM component only.  Assume that all the links are equally strong. Applying the TIM solution yields an achievable GDoF tuple $(d_{1,\mathrm{TIM}},...,d_{K,\mathrm{TIM}})$, which identifies the fraction of the signal space available to each user. Next, consider the TIN component only.  Applying appropriate power control at each transmitter and treating interference as noise at each receiver, we obtain an achievable GDoF tuple $(d_{1,\mathrm{TIN}},...,d_{K,\mathrm{TIN}})$, which identifies within the available signal space dimensions assigned to each user, the fraction of signal levels that are available to each of them. Finally, the product of the two above fractions, i.e., the GDoF tuple $(d_{1,\mathrm{TIN}}\times d_{1,\mathrm{TIM}},...,d_{K,\mathrm{TIN}}\times d_{K,\mathrm{TIM}})$, is achievable, identifying the net fraction of signal dimensions available to each user by this decomposition approach. Note that the decomposition is quite flexible, i.e., any interfering link can be considered in either TIM or TIN component (but not both simultaneously). Therefore, for one interference channel, we have multiple possible decompositions. For the TIM-TIN decomposition approach, we have the following theorem. % In general, to obtain a ``good'' achievable GDoF region, it is desirable that the TIM component contains the relatively ``strong'' interfering links and the TIN component contains the relatively ``weak'' interfering links. 

%In Fig. \ref{5user_general}, we give an example to show how to decompose an original $5$-user interference channel into a TIN component and a TIM component. 
%\begin{figure}[h]
%\centering
%\subfigure[]{
%\includegraphics[width= 4.3 cm]{5user_whole}
%\label{5user_whole}}
%\subfigure[]{
%\includegraphics[width= 4.3 cm]{5user_TIN}
%\label{5user_TIN}}
%\subfigure[]{
%\includegraphics[width= 4.3 cm]{5user_TIM}
%\label{5user_TIM}}
%\caption[]{
%\subref{5user_whole} The original 5-user interference channel, where different links may have different channel strengths; \subref{5user_TIN} one possible TIN component of the original channel; \subref{5user_TIM} The corresponding TIM component of the original channel.}
%\label{5user_general}
%\end{figure}

\begin{theorem}
\label{general_TIM_TIN_theorem}
For one specific TIM-TIN decomposition in a general $K$-user interference channel, let $\mathcal{D}_{\mathrm{TIM}}$ be the achievable GDoF region of the TIM component via signal space approach (i.e., interference alignment and ZF),  and $\mathcal{D}_{\mathrm{TIN}}$ be the achievable GDoF region of the TIN component via signal level approach (i.e., power control and TIN). Then, the following GDoF region is achievable in the original $K$-user interference channel,
\begin{align}\label{dec_proof}
\bar{\mathcal{D}}=\Big\{(d_1,&d_2,...,d_K):d_i=d_{i,\mathrm{TIM}}\times d_{i,\mathrm{TIN}},~\forall i\in[K],\nonumber\\
&\forall \mathbf{d}_{\mathrm{TIM}}=(d_{1,\mathrm{TIM}},...,d_{K,\mathrm{TIM}})\in \mathcal{D}_{\mathrm{TIM}}, ~\forall \mathbf{d}_{\mathrm{TIN}}=(d_{1,\mathrm{TIN}},...,d_{K,\mathrm{TIN}})\in \mathcal{D}_{\mathrm{TIN}}~\Big\}
\end{align}
The whole achievable GDoF region based on the TIM-TIN decomposition approach is given by
%\begin{equation}
%\label{TIM-TIN-thoerem}
$\mathcal{D}_{\mathrm{TIM-TIN}}=\mathrm{Convex~Hull}\Big(\bigcup_{\mathcal{TIM-TIN}}\bar{\mathcal{D}}\Big)$,
%\end{equation}
where $\mathcal{TIM-TIN}$ denotes the set of all the possible TIM-TIN decompositions and the convex hull operation comes from time-sharing.
\end{theorem}
%\begin{remark}
%In (\ref{TIM-TIN-thoerem}),  {\color{red}It is not hard to verify that  for a $K$-user interference channel, time-sharing between different TIM-TIN decompositions can help enlarge the achievable GDoF region in general. An example is given in Appendix ...}
%\end{remark}

\emph{Proof}: Here, the key is to prove (\ref{dec_proof}). %Given a specific TIM-TIN decomposition, assume that the GDoF tuples $\mathbf{d}_{\mathrm{TIM}}=(d_{1,\mathrm{TIM}},...,d_{K,\mathrm{TIM}})$ and $\mathbf{d}_{\mathrm{TIN}}=(d_{1,\mathrm{TIN}},...,d_{K,\mathrm{TIN}})$ are achievable in the TIM and TIN component, respectively. 
In a specific TIM-TIN decomposition, for User $k\in[K]$, denote the set of its interferers in the TIM component by $\mathcal{I}_k$. To achieve the GDoF tuple $(d_{1,\mathrm{TIM}}\times d_{1,\mathrm{TIN}},...,d_{K,\mathrm{TIM}}\times d_{K,\mathrm{TIN}})$ in the original channel, the beamforming vectors of each user are the same as those yield the GDoF tuple $\mathbf{d}_{\mathrm{TIM}}$ in the TIM component, and the power allocation for (all the data streams of) each user follows from the solution that yields the GDoF tuple $\mathbf{d}_{\mathrm{TIN}}$ in the TIN component. At the receiver, User $k$ zero-forces the interference from the users in $\mathcal{I}_k$ and treats the remaining interference as noise, which achieves the GDoF value $d_{k,\mathrm{TIM}}\times d_{k,\mathrm{TIN}}$.
%First, for the beamforming vectors, consider the TIM component only.  Assume that all the links are equally strong. Applying the TIM solution yields an achievable GDoF tuple $(d_{1,\mathrm{TIM}},d_{2,\mathrm{TIM}},...,d_{K,\mathrm{TIM}})$, which identifies the fraction of the signal space available to each user. Next, consider the TIN component only.  Applying appropriate power control at each transmitter and treating interference as noise at each receiver, we obtain an achievable GDoF tuple $(d_{1,\mathrm{TIN}},d_{2,\mathrm{TIN}},...,d_{K,\mathrm{TIN}})$, which identifies within the available signal space dimensions assigned to each user, the fraction of signal levels that are available to each of them. Finally, the product of the two above fractions, i.e., the GDoF tuple $(d_{1,\mathrm{TIN}}\times d_{1,\mathrm{TIM}},d_{2,\mathrm{TIN}}\times d_{2,\mathrm{TIM}},...,d_{K,\mathrm{TIN}}\times d_{K,\mathrm{TIM}})$, is achievable, identifying the net fraction of signal dimensions available to each user by this decomposition approach.  
\hfill $\blacksquare$

\begin{figure}[h]
\centering
\subfigure[]{
\includegraphics[width= 3.2 cm]{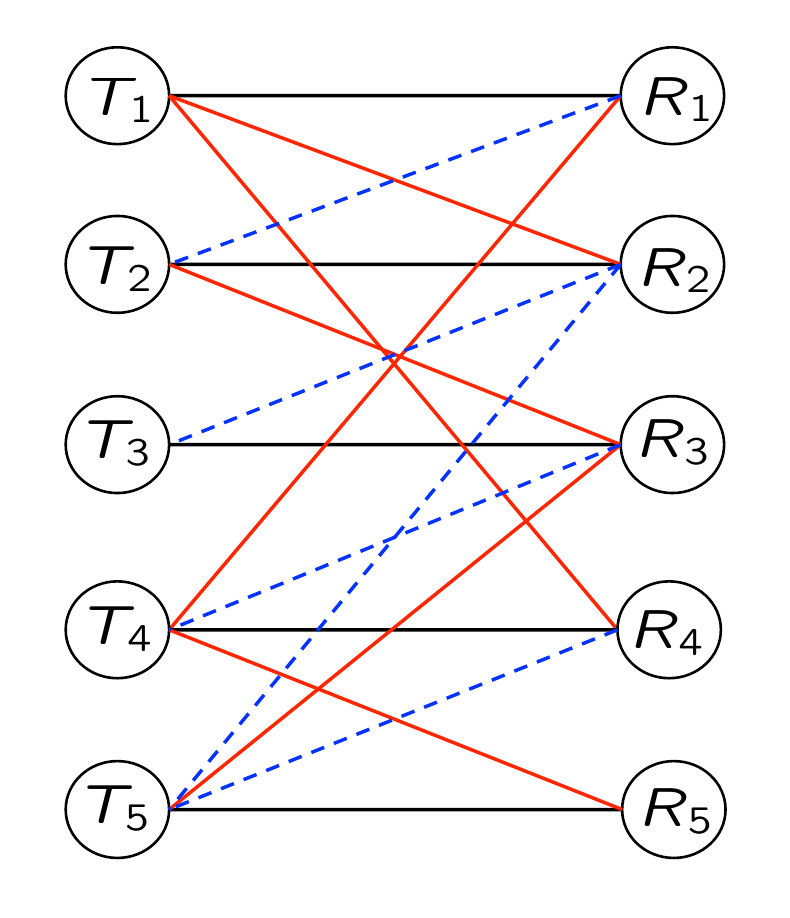}
\label{5user}}
\subfigure[]{
\includegraphics[width= 3 cm]{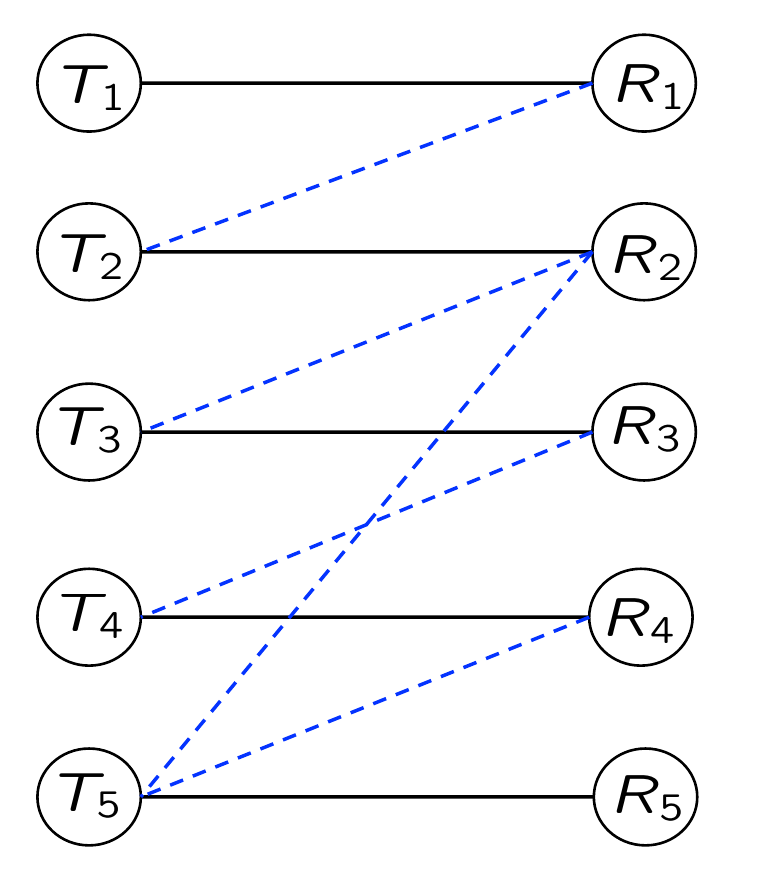}
\label{5user_M1}}
\subfigure[]{
\includegraphics[width= 3 cm]{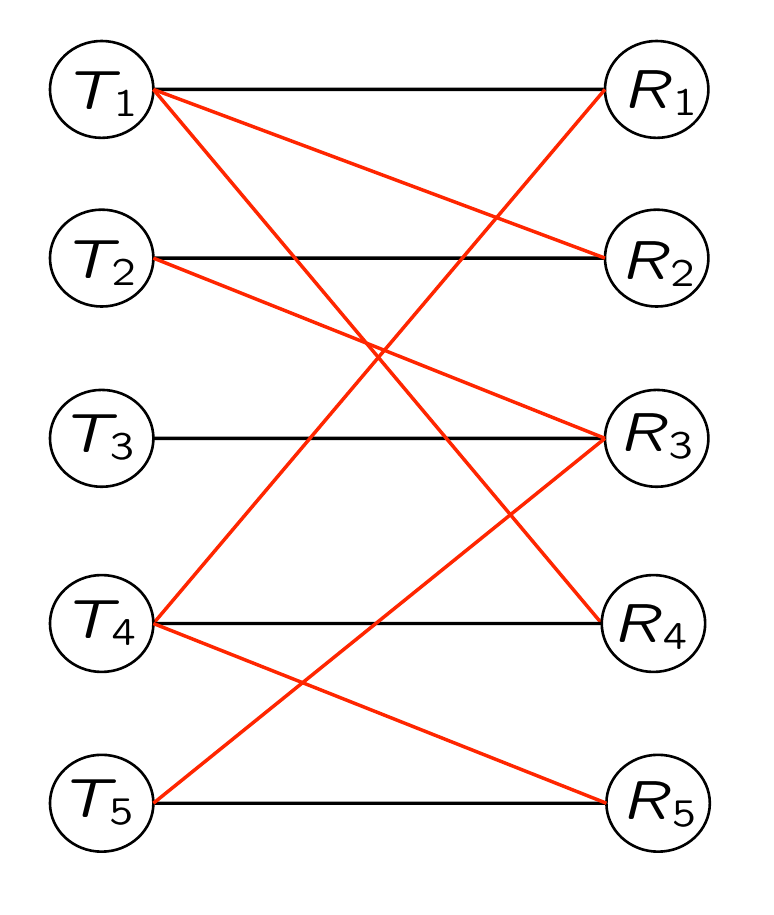}
\label{5user_S1}}
\subfigure[]{
\includegraphics[width= 4 cm]{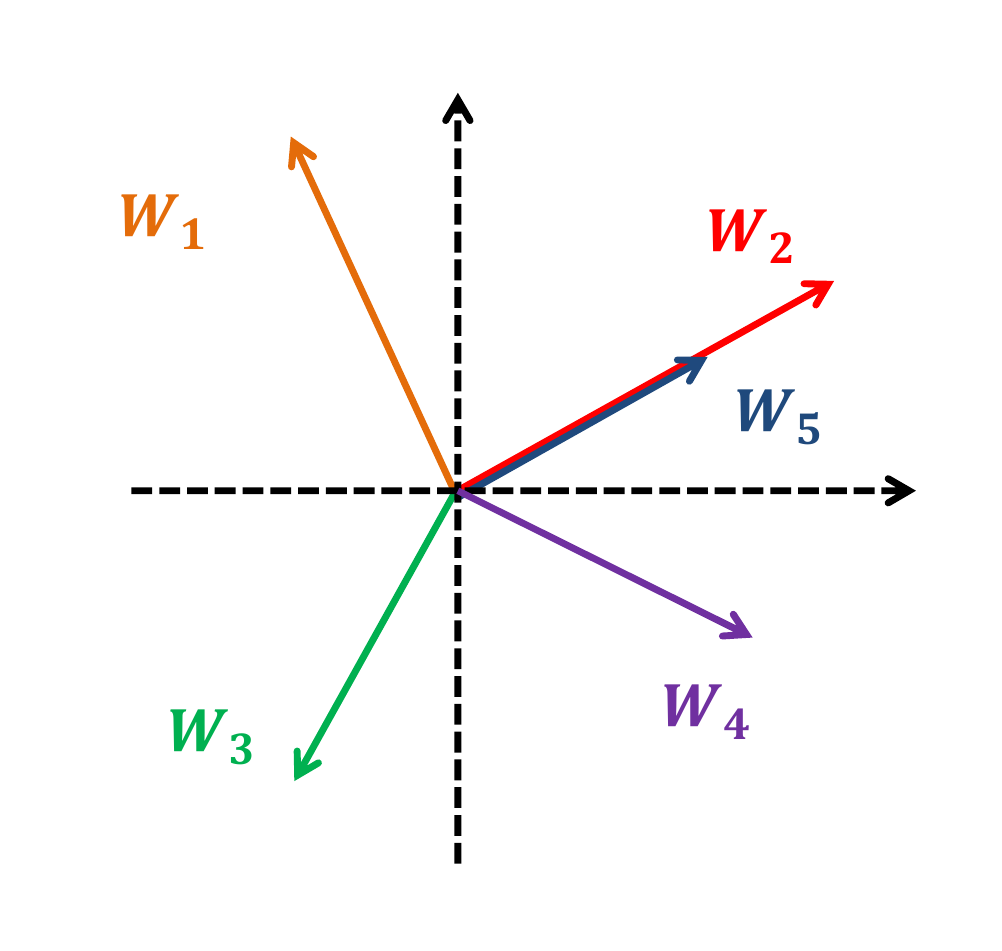}
\label{sol2}}
\caption[]{
\subref{5user} A 5-user interference channel. The red solid lines and dashed blue lines represent strong and medium interfering links, respectively. The weak interfering links are omitted to avoid cluttering the graph. \subref{5user_M1} The TIN component with all medium interfering links. \subref{5user_S1} The TIM component with all strong interfering links. \subref{sol2} The achievable scheme to achieve the symmetric GDoF value 0.3 in the original network.}
\label{5user_1}
\vspace{-0.2 in}
\end{figure}

\begin{example}\label{timtinex1}
Consider a 5-user interference channel within the QM-TIM(0,0.5) framework in Fig.~\ref{5user_1}\subref{5user}. The network is decomposed into a TIN component and a TIM component as shown in Fig.~\ref{5user_1}\subref{5user_M1} and Fig.~\ref{5user_1}\subref{5user_S1}, respectively. For the TIN component, which contains all the medium interfering links  and satisfies the TIN-optimality condition of \cite{Geng_TIN}, according to Theorem 1 in \cite{Geng_TIN} we obtain that its optimal symmetric GDoF value is $0.6$. In the TIM component, which contains all the strong interfering links, the symmetric GDoF value is $0.5$ \cite{Jafar_TIM}. Therefore, through this decomposition, in the original network the symmetric GDoF value $0.6\times0.5=0.3$ is achievable. The achievable scheme is given explicitly in Fig.~\ref{5user_1}\subref{sol2}. In this scheme, $n=2$ and $b_i=1$, $\forall i\in\{1,...,5\}$. More specifically, the achievable scheme uses a 2 dimensional space and 4 beamforming vectors, where any two of them are linearly independent and $W_2$ and $W_5$ are aligned along the same vector. The transmit power allocations are $r_1=0$, $r_2=-0.1$, $r_3=-0.2$, $r_4=-0.3$ and $r_5=-0.4$. It is easy to verify that every user achieves a GDoF value $0.3$.
\begin{itemize}
\item Receiver 1 first zero forces the interference from Transmitter 4 (to simplify notations, in the following for each Receiver $k$ we denote the interference from Transmitter $i\neq k$ by $I_i$). Then, in the remaining signal dimension, it treats the interference $I_2$ as noise. Therefore, the achievable GDoF value for Receiver $1$ is $(1-0.4)/2=0.3$.
\item Receiver 2 zero forces $I_1$ and treats $I_3$ and $I_5$ as noise to get $(0.9-0.3)/2=0.3$ GDoF.
\item Receiver 3 zero forces $I_2$ and $I_5$ and treats $I_4$ as noise to get $(0.8-0.2)/2=0.3$ GDoF.
\item Receiver 4 zero forces $I_1$ and treats $I_5$ as noise to get $(0.7-0.1)/2=0.3$ GDoF.
\item Receiver 5 zero forces $I_4$ to get $0.6/2=0.3$ GDoF.
\end{itemize}
%Note that in the achievable scheme given in Fig.~\ref{5user_1}\subref{sol2}, the transmit power allocation policy comes from the power control solution (which achieves the symmetric GDoF value 0.6) for the TIN component in Fig.~\ref{5user_M1}, and the alignment relationship for the beamforming vectors comes from the TIM solution (which achieves the symmetric GDoF value 0.5) for the TIM component in Fig.~\ref{5user_S1}.
\end{example}

As mentioned before, within the multilevel TIM framework, in general the solution based on a combination of TIM and TIN (including the decomposition approach presented in this section) is not optimal from an information theoretic perspective. However, as shown in the following, this robust decomposition approach works rather well when the quantized channel strength levels for cross links are concentrated around the bottom half of the signal levels, i.e., QM-TIM$(t_1,t_2,...,t_l)$ where $t_l\leq0.5$. For this setting, it characterizes the symmetric GDoF value to a constant factor that is no larger than 2.% (see Proposition \ref{const_factor} for details).

%Consider the $2$-user symmetric interference channel as an example, where the transmitters know the exact channel strength levels, and no phase information is available at transmitters. When the cross channel strength level is $\frac{2}{3}$, the well known ``W'' curve \cite{Tse_GDoF} shows that the Han-Kobayashi scheme achieves the optimal symmetric GDoF value $\frac{2}{3}$, which outperforms TIM-TIN decomposition.\footnote{In this case, TIM-TIN decomposition only achieves a symmetric GDoF value 1/2. The limitations of the TIM-TIN approaches include that receivers do not decode (any part of) the interference and only treat interference as \emph{white} noise.} 

%The proof for the above lemma is given in Appendix \ref{QTIM_medium_lemma_appendix}. Then we can obtain the following theorem.

\begin{theorem}
\label{const_factor}
For QM-TIM$(t_1,t_2,...,t_l)$ where $t_l\leq0.5$, the TIM-TIN decomposition approach characterizes the symmetric GDoF value $d_{\mbox{\footnotesize{sym}}}$ within a factor of $\frac{1}{1-t_l}\leq 2$. 
\end{theorem}
The proof of Theorem \ref{const_factor} is relegated to Appendix C. %\ref{appendix_prop1}

\begin{remark} The setting of QM-TIM$(t_1,t_2,...,t_l)$ where $t_l\leq0.5$ is justified by the conjecture that the optimal allocation of limited quantization bins for interfering links would be more concentrated near the noise floor. Intuitively, this is because the opportunities to communicate exist only where the desired signal significantly dominates noise/interference strengths, especially for settings with channel uncertainties where one might be forced to treat interference as noise. Although in general the optimal channel quantization is still an interesting open problem (which is beyond the scope of this paper), the above conjecture is partially settled for the $2$-user Z interference channel in \cite{Sahai_Salman_Sabharwal_Quantization}.
\end{remark}

\begin{example}
Consider the 5-user interference channel in Fig.~\ref{5user} again. In Example \ref{timtinex1}, we have shown that the symmetric GDoF value $0.3$ is achievable.  Since $0.5$ is an outer bound, the symmetric GDoF value of the original network can be characterized to a factor of $\frac{5}{3}$. In fact, we can improve this factor further. One can verify that if the medium interfering link between Transmitter $3$ and Receiver $2$ is moved from the TIN component to the TIM component, the symmetric GDoF value for the new TIN component increases to $\frac{2}{3}$ and the new TIM component still achieves the symmetric GDoF value $\frac{1}{2}$. Therefore, the achievable symmetric GDoF value via this new TIM-TIN decomposition is improved upon to $\frac{1}{3}$, and the symmetric GDoF value of the network in Fig.~\ref{5user_1}\subref{5user} is characterized to a factor of $\frac{3}{2}$. 
\end{example}

\begin{figure}[h]
\begin{center}
 \includegraphics[width= 7 cm]{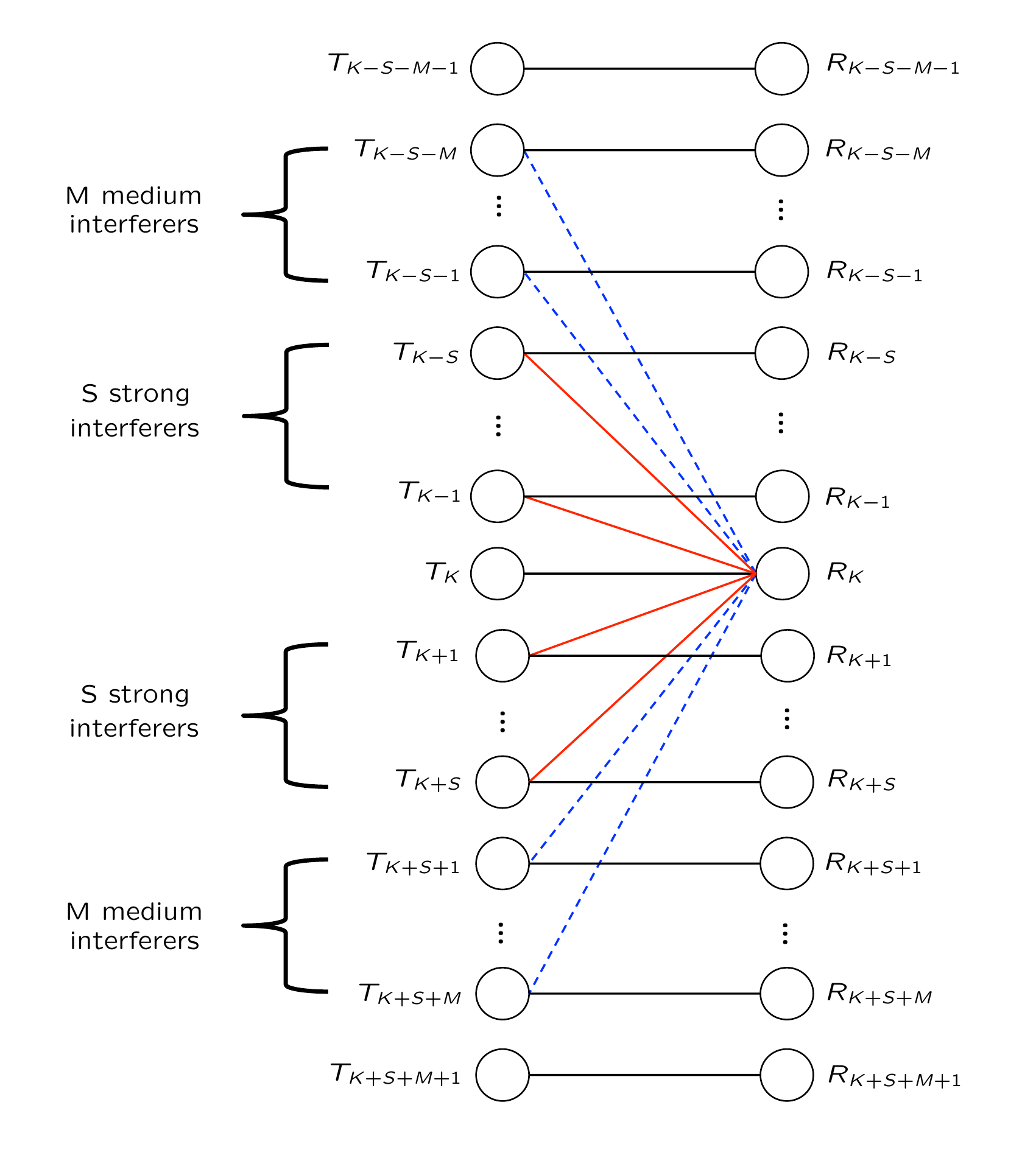}
 \caption{The symmetric multilevel neighboring interference channel with an infinite number of users. To avoid cluttering the figure,  only the direct links for users with indexes $\{K-S-M-1,...,K+S+M+1\}$ and the interring links for Receiver $k$ are shown. The red solid lines and blue dashed lines represent strong and medium interfering links, respectively.}
\label{SNIC}
\end{center}
\vspace{-0.3in}
\end{figure}

%\subsection{Optimality of TIM-TIN decomposition for multilevel neighboring interference channel}\label{neighboring_section}
Finally, we demonstrate that TIM-TIN decomposition achieves the optimal symmetric GDoF value for one broad class of multilevel neighboring interference channel, which is a natural generalization of the cellular blind interference alignment problem (or wireless index coding problem) in \cite{Maleki_Cadambe_Jafar_index}. In order to limit the number of parameters while still covering broad classes of network settings, here we mainly study symmetric cases, i.e., where relative to its own position, each receiver has the same set of strong and medium interfering links. More specifically, consider the channel depicted in Fig.~\ref{SNIC}, which is a locally connected interference channel with \emph{an infinite number} of users within the QM-TIM(0,0.5) framework. For each receiver $k$, there are $2(S+M)+1$ transmitters connected to it with channel strength level larger than the effective noise floor. One of them is the desired Transmitter $k$.  The $2S$ transmitters with indices $\{k-S,...,k-1\}$ and $\{k+1,...,k+S\}$ are connected to Receiver $k$ with strong interfering links, and the $2M$ transmitters with indices $\{k-S-M,...,k-S-1\}$ and $\{k+S+1,...,k+S+M\}$ are connected to Receiver $k$ with medium interfering links. For such networks, we have the following result.

\begin{theorem}
\label{neighboring}
For the above symmetric multilevel neighboring interference channel,  the symmetric GDoF value is
\begin{equation}
\begin{aligned}
d_{sym}=\left\{\begin{array}{cc}
            \frac{1}{S+M+1}, & M\leq S \\
            \frac{1}{2(S+1)}, & M>S
          \end{array}
          \right.
\end{aligned}
\end{equation}
which is achievable by TIM-TIN decomposition.
\end{theorem}

The proof details are provided in Appendix D. It is notable that for the symmetric neighboring interference channel, the signal space approach (with one-to-one alignments, see Appendix D) always achieves $1/(S+M+1)$ GDoF.  When $M>S+1$, according to Theorem \ref{neighboring}, the decomposition approach outperforms the pure signal space approach in terms of GDoF, and with $M$ increasing the gap between these two strictly increases.  %It is of interest to compare the performance of these two schemes in the practical SNR regime. Fig.~\ref{fig1} depicts the symmetric achievable rates of two schemes when $S=1$ and $M=4$,\footnote{In the simulation, following the channel model (\ref{equ_channel}), we keep the channel strength levels $\alpha_{ij}$ fixed and scale the parameter $P$, and we always assume that every transmitter is subject to a unit power constraint and the noise variance at each receiver is normalized to one. Since all the direct channels are with channel strength level $1$, $P$ in fact denotes the SNR of the desired link for each user.} where we find that the decomposition approach achieves a higher rate when SNR is larger than 12 dB.} 

%\begin{figure}[h]
%\begin{center}
 %\includegraphics[width= 8 cm]{fig1}
 %\caption{The achievable symmetric rates of TDMA and the decomposition approach for the symmetric multilevel neighboring interference channel where $S=1$ and $M=4$.}
%\label{fig1}
%\end{center}
%\vspace{-0.15in}
%\end{figure}

\begin{remark}
The result in Theorem \ref{neighboring} can be extended to some \textit{asymmetric} cases directly. For instance, suppose that the number of strong interferers for each user $k$ is still $2S$, whose indices are still $\{k-S,...,k-1\}$ and $\{k+1,..,k+S\}$. However, different users have different numbers of medium interferers. For User $k$, the indices of the medium interferers are $\{k-S-M_{U_k},...,k-S-1\}$ and $\{k+S+1,...,k+S+M_{D_k}\}$. If $\forall k$, $M_{U_k}>S$ and $M_{D_k}>S$, the symmetric GDoF value for such asymmetric multilevel neighboring interference channels remains as $\frac{1}{2(S+1)}$. The converse and achievability arguments both follow from the proof of Theorem \ref{neighboring}. 
\end{remark}
%Specifically, the achievable scheme is still based on TIM-TIN decomposition, in which the TIN component contains all the medium interfering links and the TIM component contains all the strong interfering links.

\section{A Distributed Numerical Approach}\label{sec_dis}
The TIM-TIN decomposition approach presented in Section \ref{sec_dcp} is a \emph{centralized analytical} method, which requires the coarse channel strength information of \emph{all} links in the network together for joint signal vector space and signal power level allocation. In this section, we devise a \emph{distributed numerical} algorithm to address the TIM-TIN problem, which only requires \emph{local} measurements on the signal strengths at each user.  The proposed algorithm is built upon a distributed power control algorithm based on the duality of TIN \cite{PC_Tin_Duality}, whose key ingredient is restated below.

\begin{lemma}\label{T1_dual} (Lemma 1 in \cite{PC_Tin_Duality})
In a general $K$-user interference channel, assume that a valid power allocation $(r_1,...,r_K)$,\footnote{In the TIN scheme, assume that the allocated power to User $i\in[K]$ is $P^{r_i}$, $r_i\leq 0$. From the GDoF perspective, we refer to the power exponent vector $(r_1,...,r_K)$ as the power allocation.} $r_i\leq 0$, $\forall i\in[K]$, achieves a GDoF tuple $(d_1,...,d_K)$. In its reciprocal channel using the power allocation $(\bar{r}_1,...,\bar{r}_K)$, where
\begin{align} \label{dual_pc}
\bar{r}_k=-\max_{j:j\neq k}\{0,\alpha_{kj}+r_j\}, ~~\forall k\in[K],
\end{align}
the achieved GDoF tuple $(\bar{d}_1,...,\bar{d}_K)$ dominates $(d_1,...,d_K)$, i.e., $\bar{d}_k\geq d_k$, $\forall k\in[K]$.
\end{lemma}

The proposed TIM-TIN distributed numerical algorithm, which is called ZEST, is specified at the top of next page.\footnote{Note that in steps 3) and 5) of the proposed ZEST algorithm, when the beam $l\in[b_k]$ of User $k\in[K]$ updates its power allocation following  Lemma \ref{T1_dual}, it treats all the remaining received beams after ZF and SC (including the other desired beams of User $k$) as interference. Also note that since in steps 2) and 4) a successive cancellation is adopted in the lexicographic order, the beam $l$ of User $k$ does not receive interference from beam s of User $k$, where $s,l\in[b_k]$ and $s<l$.} The convergence of the ZEST algorithm is given by the following theorem.

\begin{algorithm}[ht]
\caption{ZEST: {\bf ZE}ro-forcing with {\bf S}uccessive cancellation and power control for {\bf T}IM-TIN}
\begin{algorithmic}
\STATE 1) Let $m = 1$. Set $n$ and $b_k$, and randomly choose unit-norm beamforming vectors $\overrightarrow{\mathbf{v}}_{k,l}^{(m)}$ and power allocations $\overrightarrow{r}_{k,l}^{(m)}$ that satisfy the unit power constraint, $k\in[K]$, $l\in[b_k]$.
\STATE 2) In the original channel, update the receiving vectors $\overrightarrow{\mathbf{u}}_{k,l}^{(m)}$ to the unit-norm ZF-SC receiving vectors that achieve the maximal GDoF value  for each user (See Lemma \ref{lemma_ZF}. Without loss of generality, the cancellation is taken in the lexicographic order). Compute the achievable GDoF tuple $\overrightarrow{\bf{d}}^{(m)}$ and the achievable sum-GDoF value $\overrightarrow{d}_{\Sigma}^{(m)}$.
\STATE  3) Reverse the direction of the communication. Calculate the power allocation $\overleftarrow{r}_{k,l}^{(m)}$ for each data stream in the reciprocal channel following (\ref{dual_pc}), and set the beamforming and receiving vectors $\overleftarrow{\mathbf{v}}_{k,l}^{(m)}$ and $\overleftarrow{\mathbf{u}}_{k,l}^{(m)}$ as follows
\begin{align*}
%\overleftarrow{\mathbf{v}}_{k,l}^{(1)}&=\frac{\overrightarrow{\mathbf{u}}_{k,l}^{(1)}}{||\overrightarrow{\mathbf{u}}_{k,l}^{(1)}||}\\
\overleftarrow{\mathbf{v}}_{k,l}^{(m)}=\overrightarrow{\mathbf{u}}_{k,l}^{(m)},~~\overleftarrow{\mathbf{u}}_{k,l}^{(m)}=\overrightarrow{\mathbf{v}}_{k,l}^{(m)},~~ \forall k\in[K], ~\forall l\in[b_k]
\end{align*}
Compute the achievable GDoF tuple $\overleftarrow{\bf{d}}_{\mbox{\small{switch}}}^{(m)}$ and the achievable sum-GDoF value $\overleftarrow{d}_{\Sigma,\mbox{\small{switch}}}^{(m)}$ (using receivers with the \emph{reverse} lexicographic cancellation order).
\STATE 4) In the reciprocal channel, update the receiving vectors $\overleftarrow{\mathbf{u}}_{k,l}^{(m)}$ to  the unit-norm ZF-SC receiving vectors that achieve the maximal GDoF value  for each user (Again, the cancellation is taken in the lexicographic order). Compute the achievable GDoF tuple $\overleftarrow{\bf{d}}^{(m)}$ and the achievable sum-GDoF value $\overleftarrow{d}_{\Sigma}^{(m)}$.
\STATE  5) Reverse the direction of the communication. Calculate the power allocation $\overrightarrow{r}_{k,l}^{(m+1)}$ for each data stream in the original channel following (\ref{dual_pc}), and set %the beamforming and receiving vectors $\overrightarrow{\mathbf{v}}_{k,l}^{(2)}$ and $\overrightarrow{\mathbf{u}^{\star}}_{k,l}^{(2)}$ as follows
\begin{align*}
%\overrightarrow{\mathbf{v}}_{k,l}^{(2)}&=\frac{\overleftarrow{\mathbf{u}}_{k,l}^{(1)}}{||\overleftarrow{\mathbf{u}}_{k,l}^{(1)}||}\\
\overrightarrow{\mathbf{v}}_{k,l}^{(m+1)}=\overleftarrow{\mathbf{u}}_{k,l}^{(m)},~~\overrightarrow{\mathbf{u}}_{k,l}^{(m+1)}=\overleftarrow{\mathbf{v}}_{k,l}^{(m)},~~ \forall k\in[K], ~\forall l\in[b_k]
\end{align*}
Compute the achievable GDoF tuple $\overrightarrow{\bf{d}}_{\mbox{\small{switch}}}^{(m)}$ and the achievable sum-GDoF value $\overrightarrow{d}_{\Sigma,\mbox{\small{switch}}}^{(m)}$ (using receivers with the \emph{reverse} lexicographic cancellation order). Then let $m=m+1$.
\STATE 6) Repeat steps 2) through 5) until the achievable sum GDoF value (i.e., $\overrightarrow{d}_{\Sigma}^{(m)}$) converges or $m$ reaches a predefined threshold.
\end{algorithmic} 
\end{algorithm}

%\footnote{\color{blue} Similar to the GDPC algorithm, for ZF-SC-PC, the convergence to the Pareto-optimal point is not guaranteed.} 

\begin{theorem} \label{T2_convergence}
In the ZEST algorithm, $\overrightarrow{d}_{\Sigma}^{(m)}$ converges.%Numerical results for the performance of ZF-SIC-PC algorithm is given in Section \ref{s_sim}.}
\end{theorem} 
The proof of Theorem \ref{T2_convergence} is presented in Appendix E, where we show that
 \begin{align}\label{T2_result}
 &\overrightarrow{\bf{d}}^{(m)}\leq \overleftarrow{\bf{d}}_{\mbox{\small{switch}}}^{(m)}\leq  \overleftarrow{\bf{d}}^{(m)}\leq \overrightarrow{\bf{d}}_{\mbox{\small{switch}}}^{(m)} \leq \overrightarrow{\bf{d}}^{(m+1)}.%\\
 %\Rightarrow &\overrightarrow{d}_{\Sigma}^{(m)}\leq \overleftarrow{d}_{\Sigma,\mbox{switch}}^{(m)}\leq  \overleftarrow{d}_{\Sigma}^{(m)}\leq \overrightarrow{d}_{\Sigma,\mbox{switch}}^{(m)} \leq \overrightarrow{d}_{\Sigma}^{(m+1)}\nonumber
  \end{align} 
 Remarkably, the proof of Theorem \ref{T2_convergence} leads to the duality of the TIM-TIN problem naturally.  

\begin{theorem}\emph{(Duality of TIM-TIN)}
In the TIM-TIN problem, any $K$-user interference channel and its reciprocal channel have the same achievable GDoF region.
\end{theorem}
\emph{Proof}: Through (\ref{T2_result}), one can find that for any channel with arbitrary beamforming vectors and power allocations, in its reciprocal channel, we can always construct some beamforming vectors and their associated power allocations, such that the obtained GDoF tuple in the reciprocal channel dominates that achieved in the original channel. Since in the TIM-TIN problem, the achievable GDoF region for any interference channel must be upper-bounded, the original channel and its reciprocal channel have the same achievable GDoF region. \hfill $\blacksquare$

\begin{figure}[h]
\centering
\subfigure[]{
\includegraphics[width= 7.1 cm]{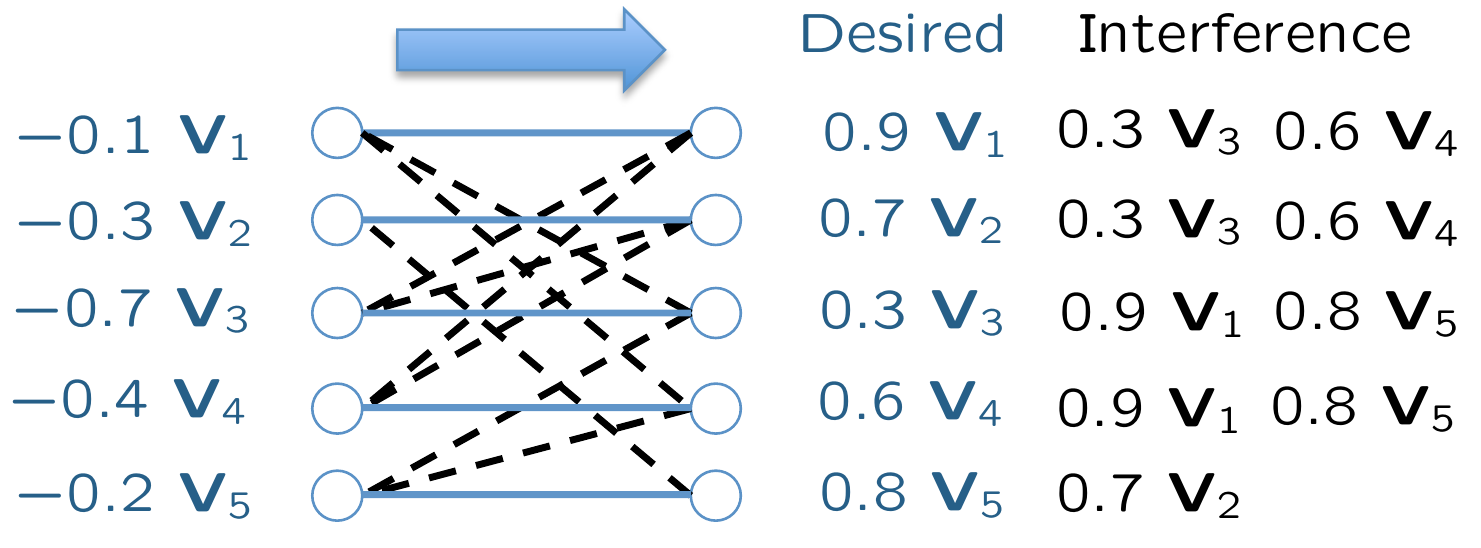}
\label{n1_ori}}
\subfigure[]{
\includegraphics[width= 7.5 cm]{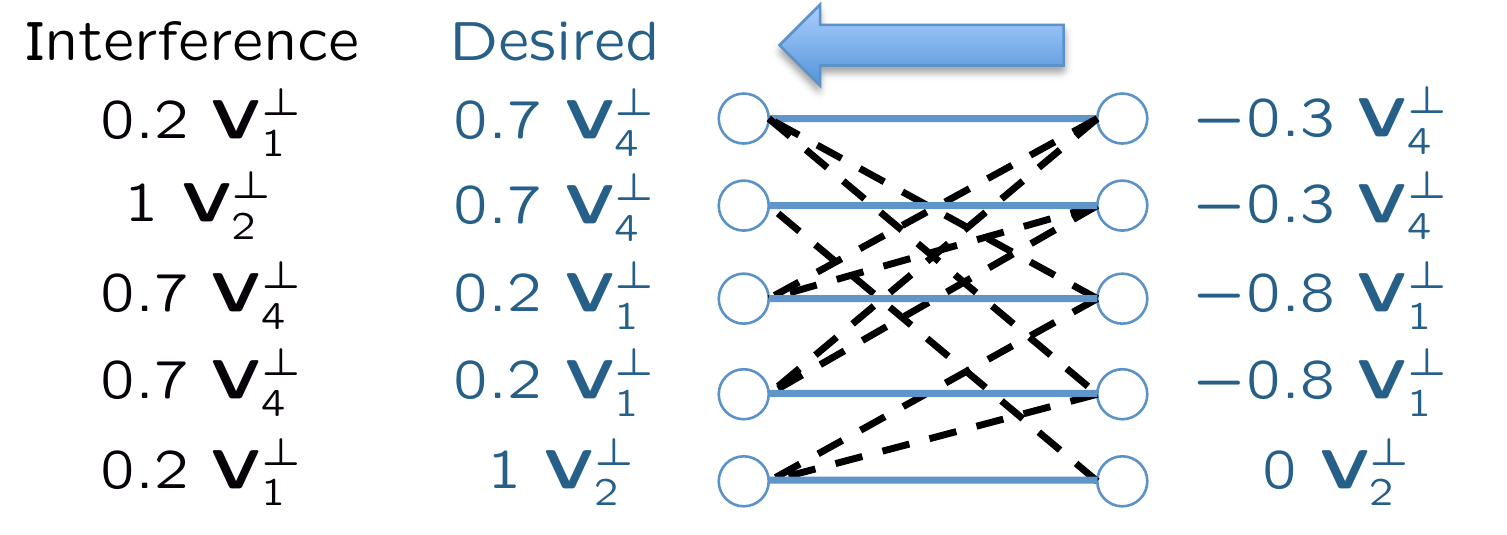}
\label{n1_rec}}
\subfigure[]{
\includegraphics[width= 6.5 cm]{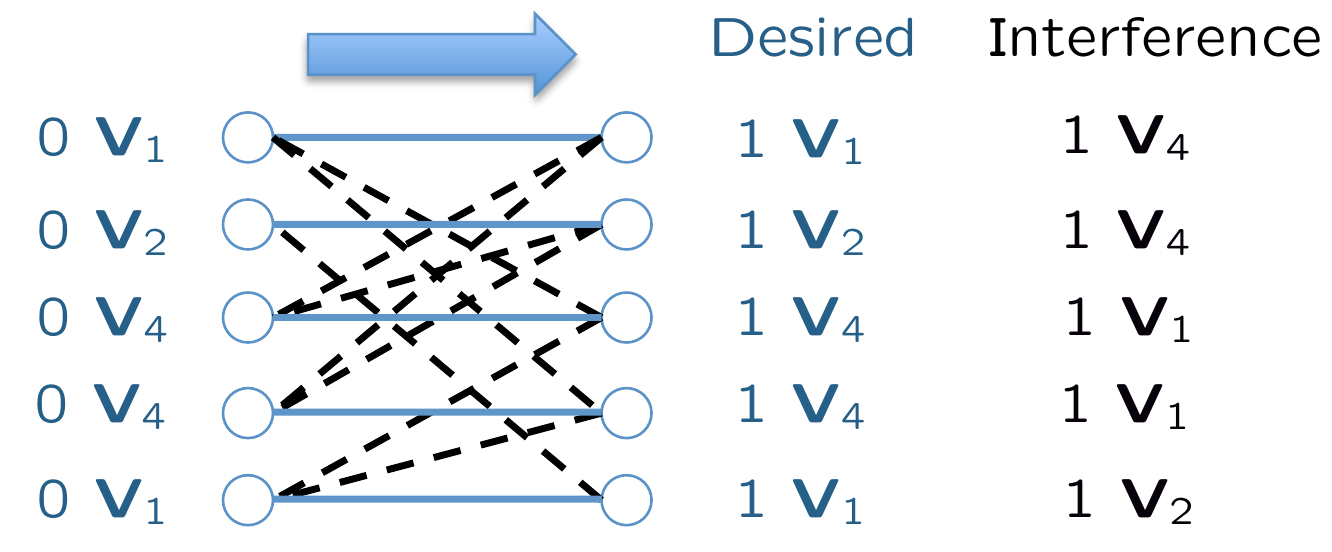}
\label{n2_ori}}
\caption[]{ 
Applying ZEST to a 5-user interference channel, where the solid blue and dash black links represent direct and cross links, respectively. In this channel, all the direct and interfering links are with channel strength level $1$, and all the other links are with channel strength level $0$ and thus omitted to avoid cluttering the graph.
\subref{n1_ori} The transmission scheme in the original channel in step 1),  \subref{n1_rec} the transmission scheme in the reciprocal channel in step 3), \subref{n2_ori} the transmission scheme in the original channel in step 5).} %\subref{n1_rec} The data transmission in the original channel when $n=2$.}
\label{n1_network}
\end{figure}

\begin{example} To help interpret how the ZEST algorithm works, consider the $5$-user interference channel in Fig.~\ref{n1_ori}. It is known that the sum-GDoF value of this channel is $2.5$ \cite{Jafar_TIM,Birk_Kol, Maleki_Cadambe_Jafar_index}. In previous literatures, the optimal solution is obtained through a centralized design. Here we show how to achieve this sum-GDoF value through the distributed ZEST algorithm.  Set $n = 2$ and $b_k=1$, $k\in\{1,...,5\}$.  Following the ZEST algorithm, in step 1), we randomly generate the beamforming vector and assign the power to each data stream for every user. As shown in Fig.~\ref{n1_ori}, the notation ``$-0.1~\mathbf{v}_1$" at the left side of Transmitter 1 denotes that the beamforming vector of User 1 is $\mathbf{v}_1$ and its allocated power level is $\overrightarrow{r}_{1,1}^{(1)} = -0.1$. All the other notations follow similarly. In step 2), each receiver updates its receiving vector. For instance, at Receiver 1, to obtain the maximal achievable GDoF value, the stronger interfering data stream with beamforming vector $\mathbf{v}_4$ is zero-forced and the weaker one is treated as noise. Therefore, we have $\overrightarrow{\mathbf{u}}_{1,1}^{(1)}=\mathbf{v}_4^{\perp}$ (where $\mathbf{v}_4^{\perp}$ denotes the $2\times 1$ vector orthogonal to $\mathbf{v}_4$), and the achievable GDoF value is $(0.9-0.3)/2=0.3$. After updating all users, the achievable GDoF tuple is $\overrightarrow{\mathbf{d}}^{(1)}=(0.3,0.2,0,0,0.4)$. Next, in step 3), we reverse the communication direction and update the power allocation for each data stream as shown in Fig.~\ref{n1_rec}. For instance, for User 1, after zero-forcing the stronger interference, the remaining interference level is 0.3. According to Lemma \ref{T1_dual}, we have $\overleftarrow{r}_{1,1}^{(1)}=-0.3$. After updating powers for all users, we obtain $\overleftarrow{\bf{d}}_{\mbox{switch}}^{(1)}=(0.35,0.35,0,0.1,0.4)$. Proceed to step 4) and update the receiving vector in the reciprocal channel. As shown in Fig.~\ref{n1_rec}, now each receiver receives one desired data stream and one interfering stream. It is easy to obtain the zero-forcing receiving vector for each user and have $\overleftarrow{\mathbf{d}}^{(1)}=(0.35,0.35,0.1,0.1,0.5)$.  Following step 5), reverse the communication direction again and update the power for each user.  At each transmitter, after zero forcing the only one interfering stream, each user sees no interference. Hence as depicted in Fig.~\ref{n2_ori}, $\overrightarrow{r}_{i,1}^{(2)}=0$, $i\in\{1,...,5\}$ and $\overrightarrow{\mathbf{d}}_{\mbox{switch}}^{(1)}=(0.5,0.5,0.5,0.5,0.5)$.  One can further verify that since then the ZEST algorithm converges, and the final solution given by this distributed algorithm is exactly the same as that obtained via the centralized design.

\end{example}

%\begin{figure}[!t]
%\centering
%\subfigure[]{
%\includegraphics[width=7 cm]{n1_ori}
%\label{n1_ori}}\\
%\subfigure[]{
%\includegraphics[width= 7 cm]{n1_rec}
%\label{n1_rec}}
%%\subfigure[]{
%%\includegraphics[width= 6.3 cm]{n2_ori}
%%\label{n2_ori}}
%\caption[]{
%\subref{n1_ori} The data transmission in the original channel when $n=1$,  where the solid blue and dash black links represent direct and cross links, respectively. All the connected links are with channel strength level $1$, and all the other links are with channel strength level $0$ and thus omitted to avoid cluttering the graph;  \subref{n1_rec} The data transmission in the reciprocal channel when $n=1$.}
%\subref{n2_ori} The data transmission in the original channel when $n=2$.}
%\vspace{-0.2in}
%\end{figure}

%\begin{figure}
 %    \centering
 %    \begin{subfigure}[b]{0.3\textwidth}
  %       \centering
   %      \includegraphics[width=\textwidth]{n1_ori}
    %     \caption{$y=x$}
     %    \label{n1_ori}
     %\end{subfigure}
     %\hfill
     %\begin{subfigure}[b]{0.3\textwidth}
      %   \centering
       %  \includegraphics[width=\textwidth]{n2_ori}
        % \caption{$y=3sinx$}
         %\label{n1_rec}
     %\end{subfigure}
       % \caption{Three simple graphs}
       % \label{fig:three graphs}
%\end{figure}

\subsection{Numerical Validations}\label{sec_num}

To further validate the GDoF performance of the proposed ZEST algorithm, we consider a random $5$-user interference channel. We assume that the channel strength levels of all direct links are always equal to $1$. Motivated by cellular networks where users suffer strong interference from neighboring cells, we assume that at Receiver $i$, the interference from Transmitter $i-1$ and $i+1$ are strong interference, and the others are weak.\footnote{Here we consider a cyclic setting, i.e., when $i=1$, $i-1=5$ and when $i=5$, $i+1=1$.} For the strong interference, we assume that their  channel strength levels fall into a uniform distribution of $[x, 1]$, and the channel strength levels of the weak interfering links  fall into a uniform distribution of $[0, 1-x]$, where $x\geq 0.5$.
Following \cite{Geng_PC, Yi_PC, PC_Tin_Duality}, we keep the channel strength levels $\alpha_{ij}$ fixed and scale the parameter $P$ in each random channel realization, and we always assume that every transmitter is subject to a unit peak power constraint and the noise variance at each receiver is normalized to one. Since all the direct channels are with channel strength level $1$, $P$ in fact denotes the SNR of the desired link for each user.  

We compare the achievable sum-GDoF of the proposed ZEST algorithm, the well-known distributed interference alignment algorithm Max-SINR \cite{Gomadam_DIA},\footnote{The Max-SINR algorithm is originally proposed for MIMO interference channels. Here we adopt the algorithm for SISO interference channels with multiple channel uses.} the state-of-the-art power control algorithm SAPC (i.e., SINR approximation power control) \cite{Chiang_DPC}, TDMA (i.e., the orthogonal scheme with equal time sharing among all users) and the full power transmission (i.e., every user always utilizes full power to transmit its own signal). It is notable that Max-SINR and SAPC optimizes the signal space allocation and signal level allocation, respectively. Among all the schemes considered here, only ZEST jointly optimizes the signal space and signal level allocation for data transmission.   For both ZEST and Max-SINR, we set the number of channel uses $n$ as 2 and the number of scalar data streams for each user $d_k$ as 1, $\forall k\in\{1,...,5\}$. We also note that for both ZEST and Max-SINR, different initializations may yield different sum-rates, particularly for ZEST in low and medium SNR regimes.\footnote{We point out that the convergence of  Max-SINR is still open. In our experiment, we note that for Max-SINR, with a sufficient number of iterations, different initializations usually converge to the same sum-rate. In practice, when the number of iterations is limited, different initializations may lead to different final solutions though.  In \cite{Wilson_Cnvg_MSINR} a convergent Max-SINR algorithm is developed, which in fact jointly optimizes the signal vector space and signal power level allocations. However, the proposed algorithm in \cite{Wilson_Cnvg_MSINR} is based on the duality of SINR in multiuser MIMO networks under an artificial \emph{sum power constraint} \cite{Song_Duality}.  While in ZEST, the convergence is guaranteed under the practical \emph{individual user power constraint}. But due to the non-convexity of the problem, the convergent point depends on the initialization. For SAPC, following \cite{Chiang_DPC}  we always set the initial power of each user as its maximal transmit power.}  In our experiment, for both ZEST and Max-SINR, in each channel realization we start from multiple random initializations and pick the largest yielded sum-rate as the final solution. When the SNR value $P$ is less than 30 dB, we set the number of random initializations as 30, and 10 otherwise. How to smartly choose the initialization of ZEST to improve sum-rate in low and medium SNR regimes is an interesting open question. 
 
\begin{figure}[h]
\centering
\subfigure[]{
\includegraphics[width= 7 cm]{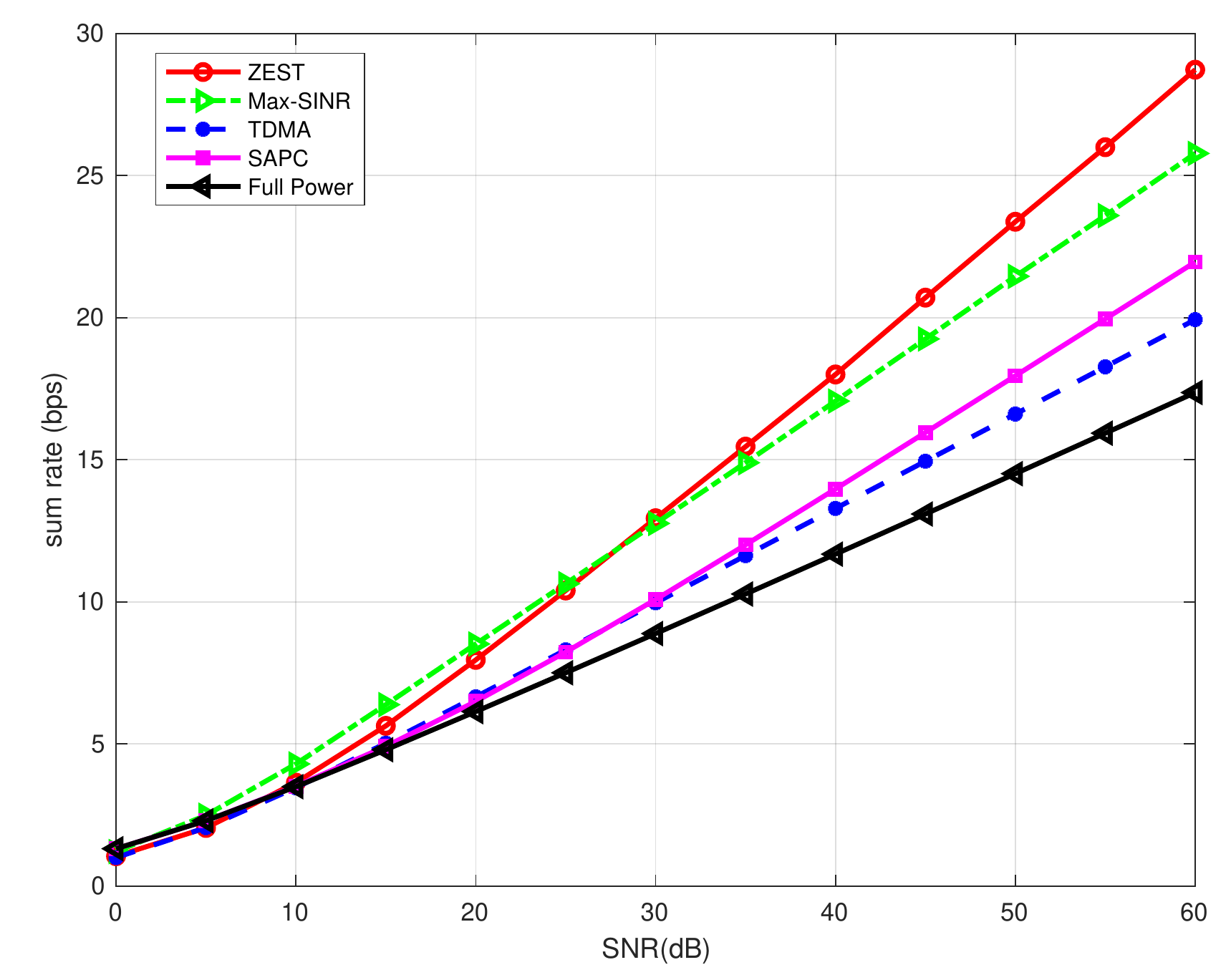}
\label{30_10_seeds_200sims_peak}}
\subfigure[]{
\includegraphics[width= 7 cm]{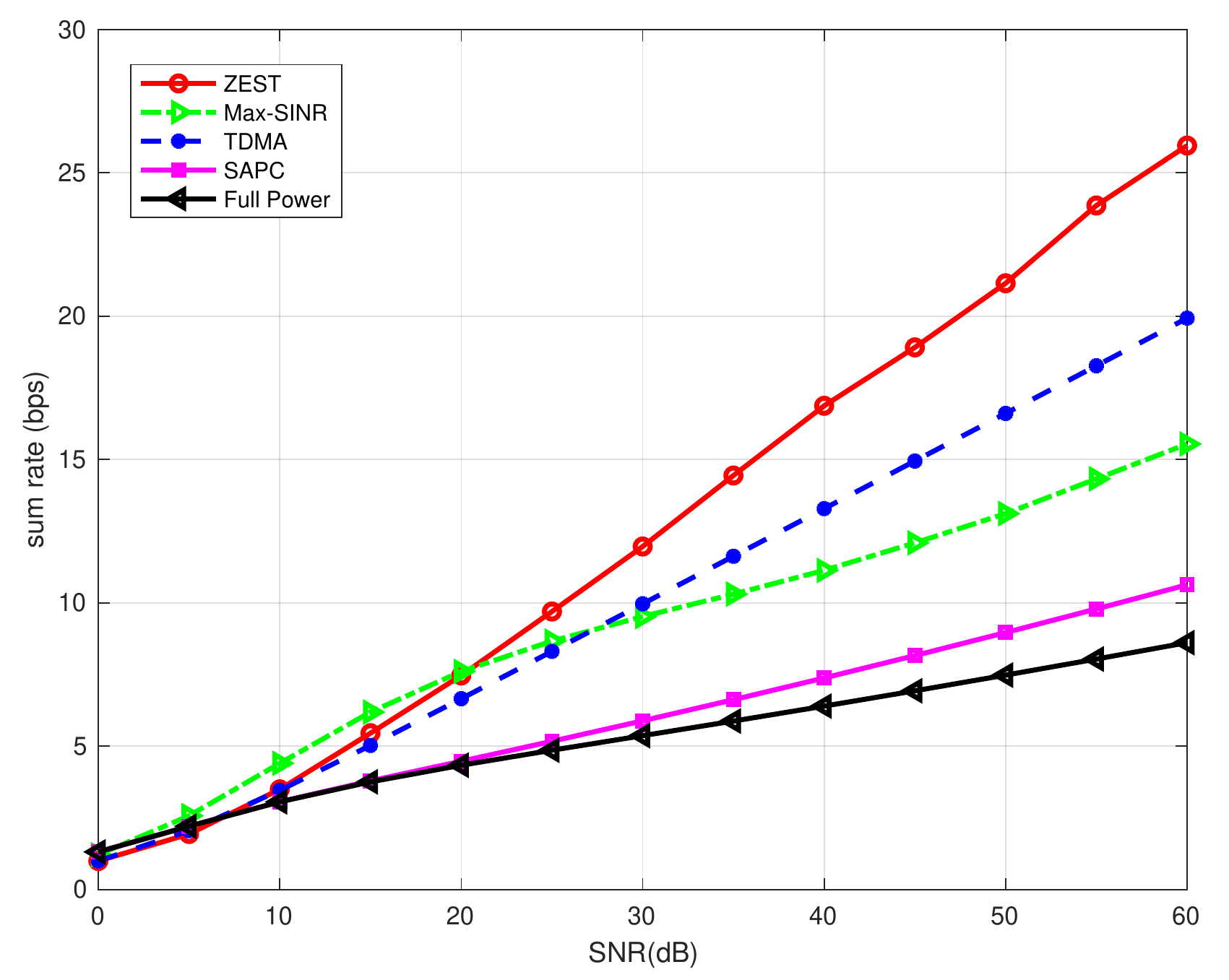}
\label{30_10_seeds_200sims_peak_set2}}
\caption[]{ 
Sum-rate performance of ZEST, Max-SINR, TDMA, SAPC, and the full power transmission, when \subref{30_10_seeds_200sims_peak} $x=0.5$, and  \subref{30_10_seeds_200sims_peak_set2} $x=0.75$, where the latter models the settings with more diverse channel strengths between strong and weak interfering links.} 
\label{num_sim}
\vspace{-0.1in}
\end{figure}

In our experiments, we consider two specific $x$ values, i.e., $x=0.5$ and $x=0.75$, where the latter models the settings with more diverse channel strengths between strong and weak interfering links. For the two different $x$ values, the averaged sum-rate of all algorithms over 200 random channel realizations are given in Fig.~\ref{30_10_seeds_200sims_peak} and \ref{30_10_seeds_200sims_peak_set2}, respectively. It can be seen that in both cases ZEST achieves the largest sum-GDoF value (i.e., the steepest slope in the high SNR regime) among all the schemes. More interestingly, ZEST outperforms SAPC, TDMA and the full power transmission almost over the entire SNR range. Compared with Max-SINR, ZEST is particularly favorable in the settings with more disparate interference strengths (e.g., when $x=0.75$), and in both cases ZEST only suffers slight sum-rate degradation when the SNR value is relatively low. 

%Comparing the slopes of the sum rates in the high SNR regime, it is not hard to obtain that in terms of GDoF, in the considered scenario the Max-SINR algorithm only achieves comparable performance with the TDMA scheme, and the ZF-SIC-PC algorithm achieves a sum-GDoF gain of about $30\%$ compared with Max-SINR and TDMA. As the figure shows, this GDoF gain will translate to an unbounded rate gain in high SNR limits. In the relatively low SNR regime, the ZF-SIC-PC algorithm suffers a small rate penalty due to the naive ZF procedure and a coarse GDoF based power control. How to improve the achievable rate in the relatively low SNR setting by jointly optimizing signal space and signal level allocations is still open.
 
\begin{figure}[h]
\begin{center}
 \includegraphics[width=7 cm]{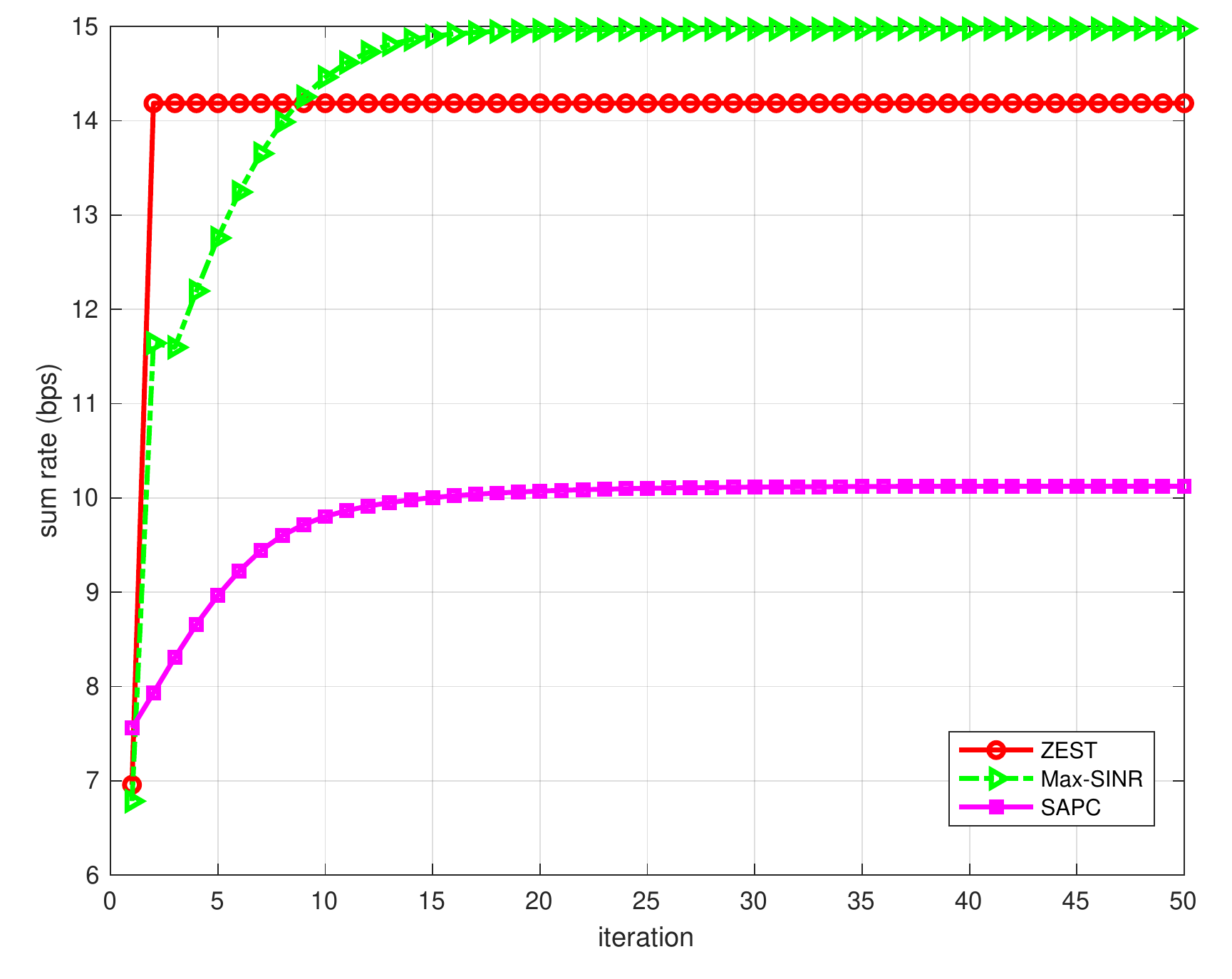}
 \caption{A representative example for the convergence behavior of ZEST, max-SINR, and SAPC.}
\label{conv_30dB_2}
\end{center}
\vspace{-0.2in}
\end{figure}

Next, we consider the convergence of the ZEST algorithm. In general, the numerical results show that in all channel realizations and in all SNR regimes, ZEST exhibits a much faster convergence rate than Max-SINR and SAPC. In our experiment, a few iterations are usually sufficient for ZEST's convergence. A representative example is given in Fig.~\ref{conv_30dB_2} when $x=0.5$ and SNR = 30 dB. Note that as shown in Fig.~\ref{conv_30dB_2}, the convergence of Max-SINR is not always monotone, which has been reported in \cite{Wilson_Cnvg_MSINR} as well.

\subsection{Discussions}

As noted in Section \ref{sec_num}, the GDoF-based algorithm ZEST in general converges much faster than Max-SINR and SAPC, which are  optimization algorithms both based on the classical SINR metric. We have similar observations for the GDoF-based power control algorithm iGPC (iterative GDoF-duality-based power control) \cite{PC_Tin_Duality}, which usually exhibits a much faster convergence rate than SAPC.  An interesting question one may ask is if the outputs of the GDoF-based algorithms, which converge faster, could be used as initializations to speed up the convergence of their conventional counterparts.  

\begin{figure}[h]
\centering
\subfigure[]{
\includegraphics[width= 7 cm]{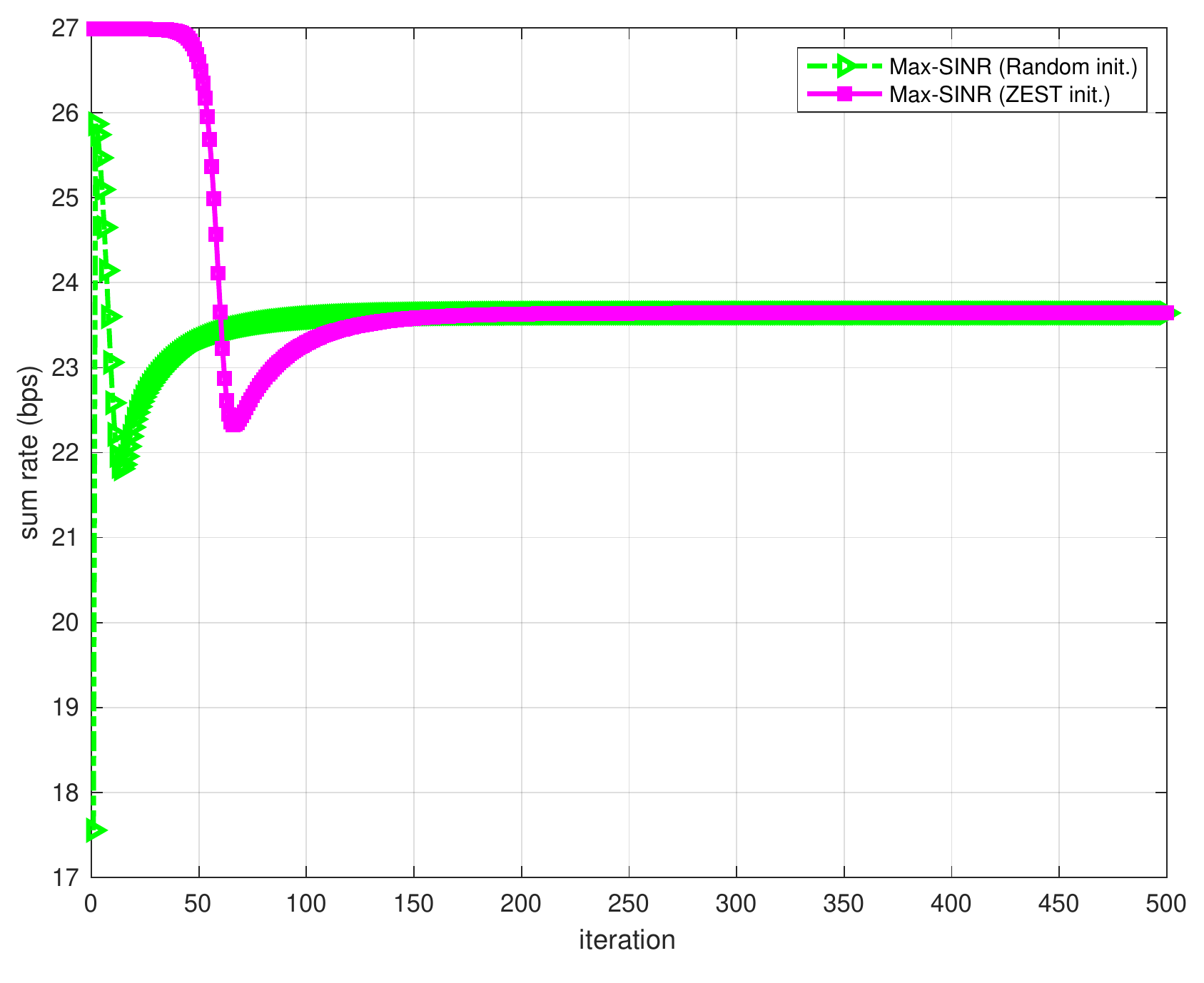}
\label{ms_conv2}}
\subfigure[]{
\includegraphics[width= 7 cm]{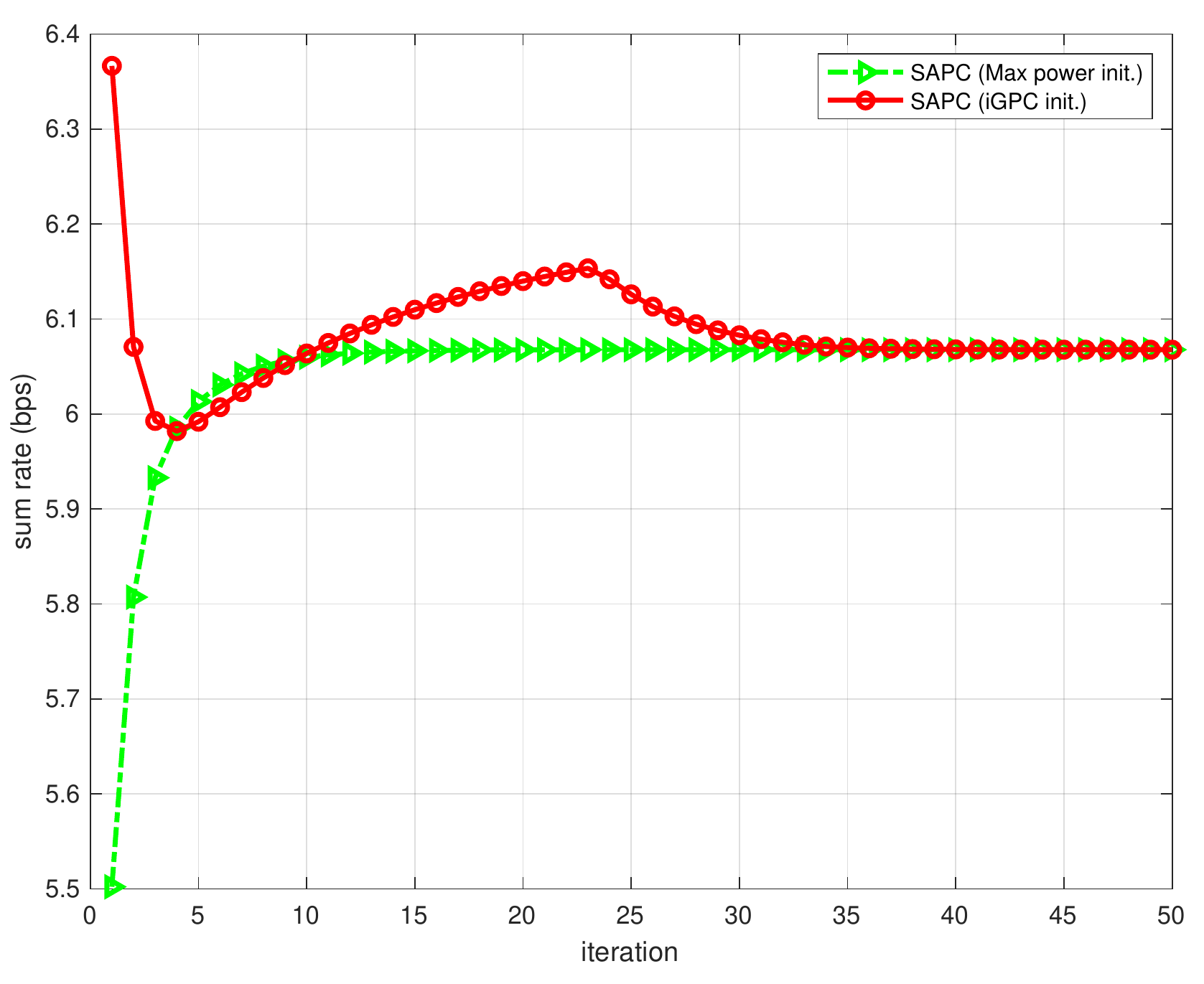}
\label{SAPC_conv1}}
\caption[]{ 
Examples where using the GDoF-based solutions as initializations does not speed up the convergence of conventional algorithms: \subref{ms_conv2}  Max-SINR and \subref{SAPC_conv1} SAPC, as these two algorithms do not always converge monotonically.}
\label{conv_sim}
\vspace{-0.1in}
\end{figure}

Here, we consider using the output beamforming vectors of ZEST as the initialization of Max-SINR, and the output power allocations of iGPC as the initialization of SAPC.  We note that in our experiment, compared with conventional initialization methods (i.e., random initialization in Max-SINR, and maximum power initialization in SAPC \cite{Chiang_DPC}), using the GDoF-based solution as a starting point usually means starting from a higher sum-rate, and thus helps speed up the optimizations in Max-SINR and SAPC in many cases.  However, the answer to the question asked above is not always positive. Two counter examples are given in Fig.~\ref{conv_sim}. The main observation here is that neither Max-SINR nor SAPC is guaranteed to converge monotonically. Therefore, starting from a higher sum rate does not always lead to faster convergence for these two algorithms. 

\section{Conclusion}
In this paper, we formulate a joint signal vector space and signal power level allocation problem (i.e., the TIM-TIN problem) under the assumption that only a coarse knowledge of channel strengths and no knowledge of channel phases is available to the transmitters. A decomposition of the problem into TIN and TIM components is proposed as a baseline.  A distributed numerical algorithm called ZEST is developed as well. The convergence of the ZEST algorithm leads to the duality of the TIM-TIN problem. The joint TIM-TIN approach is promising to be a fundamental building block in existing and future wireless networks, due to its robustness to channel knowledge at transmitters, implementation simplicity (e.g., no need to decode any interference, and being implemented in a distributed fashion) and potential superior performance. This line of research is still in its infancy though. It is hoped that this work could inspire more future research in this area. Future direction include, e.g., translating theoretical insights obtained in this work into the design of practical large-scale wireless networks, such as  device-to-device networks and heterogeneous cellular networks.

\section*{Appendix A: Proof of Lemma \ref{Matrix_lemma}}
%\section{Proof of Lemma \ref{Matrix_lemma}} \label{Matrix_lemma_proof}
Let $x_i\sim \mathcal{CN}(0,P^{\kappa_i})$ be independent Gaussian variables. Denote by $\mathbf{z}$ an $n\times 1$ zero mean unit variance circularly symmetric Gaussian vector. When $m>\gamma$, denote the $n\times1$ vectors $\mathbf{v}_i\nsubseteq\mathcal{V}_{\Pi}$ as $\mathbf{v}_{\Pi'(j)}$, $j\in[m-\gamma]$.  We have
\begin{align}
&\log\Big[\det\Big(\mathbf{I}+\sum_{i=1}^mP^{\kappa_i}\mathbf{v}_i\mathbf{v}_i^\dag\Big)\Big]\nonumber\\
=&~h\Big(\sum_{i=1}^m\mathbf{v}_ix_i+\mathbf{z}\Big)+o(\log(P))\\
=&~h\Big(\sum_{i=1}^{\gamma}\mathbf{v}_{\Pi(i)}x_{\Pi(i)}+\sum_{j=1}^{m-\gamma}\mathbf{v}_{\Pi'(j)}x_{\Pi'(j)}+\mathbf{z}\Big)+o(\log(P))\\\label{simplify}
=&~h\Big(\sum_{i=1}^{\gamma}\mathbf{v}_{\Pi(i)}x_{\Pi(i)}+\mathbf{z}\Big)+o(\log(P))\\
=&~\log\Big[\det\Big(\mathbf{I}+\sum_{i=1}^{\gamma}P^{\kappa_{\Pi(i)}}\mathbf{v}_{\Pi(i)}\mathbf{v}_{\Pi(i)}^\dag\Big)\Big]+o(\log(P)),
\end{align}
where (\ref{simplify}) is due to the facts that $\mathbf{v}_{\Pi'(j)}$, $\forall j\in[m-\gamma]$ is a linear combination of the vectors in $\mathcal{V}_{\Pi} =\{\mathbf{v}_{\Pi(1)}, \mathbf{v}_{\Pi(2)},...,\mathbf{v}_{\Pi(\gamma)}\}$, and the term $\sum_{j=1}^{m-\gamma}\mathbf{v}_{\Pi'(j)}x_{\Pi'(j)}$ becomes insignificant when $P$ approaches infinity. More specifically, as $P\rightarrow \infty$, for the term $\mathbf{v}_i(x_i+x_j+...+x_k)$ ($i<j<...<k$), only the symbol $x_i$ with the dominant power exponent $\kappa_i$ matters, implying that for the vector $\mathbf{v}_i$ we can ignore all the other independent symbols with equal or smaller power exponents in the limit of $P\rightarrow\infty$. 

The following is essentially the same as the proof of Lemma 1 in \cite{Gou_Jafar_SIMO}. Define $\mathbf{V}_{\Pi}\triangleq[\mathbf{v}_{\Pi(1)}~\mathbf{v}_{\Pi(2)}~...~\mathbf{v}_{\Pi(\gamma)}~]$ with size $n\times \gamma$, and the diagonal matrix $\mathbf{P}_{\Pi}\triangleq\mbox{diag}[P^{\kappa_{\Pi(1)}}~P^{\kappa_{\Pi(2)}}~...~P^{\kappa_{\Pi(\gamma)}}]$ with size $\gamma\times \gamma$. We have
\begin{align}
&\log\Big[\det(\mathbf{I}+\sum_{i=1}^{\gamma}P^{\kappa_{\Pi(i)}}\mathbf{v}_{\Pi(i)}\mathbf{v}_{\Pi(i)}^\dag)\Big]\nonumber\\
=&\log\Big[\det(\mathbf{I}+\mathbf{V}_{\Pi}\mathbf{P}_{\Pi}\mathbf{V}_{\Pi}^{\dag})\Big]\\
=&\log\Big[\det(\mathbf{I}+\mathbf{V}_{\Pi}^{\dag}\mathbf{V}_{\Pi}\mathbf{P}_{\Pi})\Big]\\
=&\log\Big[\det(\mathbf{P}_{\Pi})\Big]+\log\Big[\det(\mathbf{P}_{\Pi}^{-1}+\mathbf{V}_{\Pi}^{\dag}\mathbf{V}_{\Pi})\Big]\\
=&\sum_{i=1}^\gamma \kappa_{\Pi(i)}\log P +\mathcal{O}(1)
\end{align}

\section*{Appendix B: Proof of Lemma \ref{lemma_ZF}}
%\section{Proof of Lemma \ref{lemma_ZF}}\label{app_ZF}
Recall  that in Section \ref{sec_timtin_form}, from vectors $\mathcal{V}=\{\mathbf{v}_1,...,\mathbf{v}_m\}$ and their associated power exponents $\mathcal{R}=\{\mathbf{\kappa}_1,...,\mathbf{\kappa}_m\}$, 
we obtain $\gamma\leq n$ linearly independent beamforming vectors $\mathcal{V}_{\Pi} =\{\mathbf{v}_{\Pi(1)}, \mathbf{v}_{\Pi(2)},...,\mathbf{v}_{\Pi(\gamma)}\}$ and their associated power exponents $\mathcal{P}_{\Pi}=\{ \kappa_{\Pi(1)},\kappa_{\Pi(2)},...,\kappa_{\Pi(\gamma)}\}$. 
Define these operations as $\mathcal{N}_v$ and $\mathcal{N}_{\kappa}$, respectively, i.e., $\mathcal{N}_v(\mathcal{V},\mathcal{R})=\mathcal{V}_{\Pi}$ and $\mathcal{N}_{\kappa}(\mathcal{V},\mathcal{R})=\mathcal{P}_{\Pi}$. Denote by $\kappa_{\Sigma,\mathcal{N}_{\kappa}(\mathcal{V},\mathcal{R})}$ the sum of all entries in $\mathcal{N}_{\kappa}(\mathcal{V},\mathcal{R})$.

To prove lemma \ref{lemma_ZF}, without loss of generality, we only need to consider User $1$ and assume that the successive cancellation is taken in the lexicographic order. According to the chain rule, we have
\begin{align}
R_1=\frac{1}{n}I(s_{1,1},s_{1,2},...,s_{1,b_1};\mathbf{y}_1)=\sum_{i=1}^{b_1} \underbrace{\frac{1}{n}I(s_{1,i};\mathbf{y}_1|s_{1,1},...,s_{1,i-1})}_{\triangleq R_{1,i}}
\end{align}
Let $d_{1,i}=\lim_{P\rightarrow \infty}\frac{R_{1,i}}{\log P},~~\forall i\in[b_1]$. We have
\begin{align}
d_1=\sum_{i=1}^{b_1}d_{1,i}\label{chain}
\end{align}

For Receiver $1$, denote the sets of the beamforming vectors and associated power exponents for all the received data streams as $\mathcal{V}_1$ and $\mathcal{R}_1$, respectively. Consider each term in the right hand side of (\ref{chain}). Start with $d_{1,1}$. We have the following  two cases.
\begin{itemize}
\item $r_{1,1}+\alpha_{11}\in \mathcal{N}_{\kappa}(\mathcal{V}_1,\mathcal{R}_1)$: In this case,  we have $\mathbf{v}_{1,1}\in\mathcal{N}_v(\mathcal{V}_1,\mathcal{R}_1)$. From Lemma \ref{Matrix_lemma}, we have
\begin{align*}
R_{1,1}=\frac{1}{n}\Big[h(\mathbf{y}_1)-h(\mathbf{y}_1|s_{1,1})\Big]=\frac{1}{n}\Big[\kappa_{\Sigma,\mathcal{N}_{\kappa}(\mathcal{V}_1,\mathcal{R}_1)}-\kappa_{\Sigma,\mathcal{N}_{\kappa}(\mathcal{V}_1\backslash
\mathbf{v}_{1,1},\mathcal{R}_1\backslash \{r_{1,1}+\alpha_{11}\})} \Big]\log P+o(\log(P))
\end{align*}
Therefore, in the GDoF sense, we have
%\begin{align*}
$d_{1,1}=\frac{\kappa_{\Sigma,\mathcal{N}_{\kappa}(\mathcal{V}_1,\mathcal{R}_1)}-\kappa_{\Sigma,\mathcal{N}_{\kappa}(\mathcal{V}_1\backslash
\mathbf{v}_{1,1},\mathcal{R}_1\backslash\{ r_{1,1}+\alpha_{11}\})}}{n}$,
%\end{align*}
which is achievable by zero-forcing all the data streams falling into $\mathrm{span}(\mathcal{N}_v(\mathcal{V}_1,\mathcal{R}_1)\backslash \mathbf{v}_{1,1})$ and treating the remaining interference as noise.

 \item $r_{1,1}+\alpha_{11}\notin \mathcal{N}_{\kappa}(\mathcal{V}_1,\mathcal{R}_1)$: In this case,  $\mathbf{v}_{1,1}\notin\mathcal{N}_v(\mathcal{V}_1,\mathcal{R}_1)$. We have $R_{1,1}=o(\log(P))$ and $d_{1,1}=0$, which is trivially achievable (by ZF and TIN).
\end{itemize}
After decoding $s_{1,1}$, we subtract it out from the received signal and then consider the second term in the right hand side of (\ref{chain}), i.e., $d_{1,2}$. Similarly, we can argue that by zero-forcing certain interfering data streams for $s_{1,2}$ and treating others as noise, $d_{1,2}$ is achievable. Repeating this subtract-and-decode argument until all the desired data streams for User $1$ are decoded,  we establish that $d_1$ is achievable via the ZF-SC receiver and complete the proof.

\section*{Appendix C: Proof of Theorem \ref{const_factor}} \label{appendix_prop1}
%According to Theorem \ref{general_TIM_TIN_theorem} and the fact that removing interfering links in interference channels does not hurt capacity, we have

%where $d_{\mbox{\footnotesize{sym}}}^{\mbox{\footnotesize{TIN}}}$ is the symmetric GDoF value of the TIN component containing all the interfering links with channel strength levels no stronger than $t_l$, and $d_{\mbox{\footnotesize{sym}}}^{\mbox{\footnotesize{TIM}}}$ is the symmetric GDoF value of the TIM component containing all the interfering links with channel strength levels stronger than $t_l$.  %And $d_{\mbox{\footnotesize{sym}}}$ is characterized to a factor no larger than $\frac{1}{1-t_l}$.

In the achievability, the original network is decomposed into a TIN component containing all the interfering links with channel strength levels no stronger than $t_l$ and a TIM component containing all the other interfering links. First, consider the TIN component. When $t_l\leq0.5$, the TIN component satisfies the TIN-optimality condition identified in \cite{Geng_TIN} (recall  that the channel strength level of the direct link is normalized to 1). Following Theorem 1 in \cite{Geng_TIN}, its symmetric GDoF value $d_{\mbox{\footnotesize{sym}}}^{\mbox{\footnotesize{TIN}}}\geq1-t_l$. Next, for the TIM component, assume that given the optimal signal space solution, the optimal symmetric GDoF value is denoted by $d_{\mbox{\footnotesize{sym}}}^{\mbox{\footnotesize{TIM}}}$. Finally, according to Theorem \ref{general_TIM_TIN_theorem} in this paper, the symmetric GDoF value $d_{\mbox{\footnotesize{sym}}}^{\mbox{\footnotesize{TIM}}}\times d_{\mbox{\footnotesize{sym}}}^{\mbox{\footnotesize{TIN}}}$ is achievable. 

For the converse, $d_{\mbox{\footnotesize{sym}}}^{\mbox{\footnotesize{TIM}}}$ and $d_{\mbox{\footnotesize{sym}}}^{\mbox{\footnotesize{TIN}}}$ are both outer bounds for the original network, since removing interfering links from the channel does not decrease GDoF. Therefore, $\min\{d_{\mbox{\footnotesize{sym}}}^{\mbox{\footnotesize{TIM}}},d_{\mbox{\footnotesize{sym}}}^{\mbox{\footnotesize{TIN}}}\}$ can serve as an outer bound for the symmetric GDoF value of the original network. We have
%\begin{align}
%\label{dsymA}
$d_{\mbox{\footnotesize{sym}}}^{\mbox{\footnotesize{TIN}}}\times d_{\mbox{\footnotesize{sym}}}^{\mbox{\footnotesize{TIM}}}\leq d_{\mbox{\footnotesize{sym}}}\leq \min\{d_{\mbox{\footnotesize{sym}}}^{\mbox{\footnotesize{TIN}}},d_{\mbox{\footnotesize{sym}}}^{\mbox{\footnotesize{TIM}}}\}$,
%\end{align}
and the symmetric GDoF value $d_{\mathrm{sym}}$ can be characterized to a factor
\begin{align}
\beta&=\frac{\min\{d_{\mbox{\footnotesize{sym}}}^{\mbox{\footnotesize{TIM}}},d_{\mbox{\footnotesize{sym}}}^{\mbox{\footnotesize{TIN}}}\}}{d_{\mbox{\footnotesize{sym}}}^{\mbox{\footnotesize{TIN}}}\times d_{\mbox{\footnotesize{sym}}}^{\mbox{\footnotesize{TIM}}} }\leq \frac{\min\{d_{\mbox{\footnotesize{sym}}}^{\mbox{\footnotesize{TIM}}},1\}}{(1-t_l)\times d_{\mbox{\footnotesize{sym}}}^{\mbox{\footnotesize{TIM}}}}
%&=\min\{\frac{1}{1-t_l},\frac{1}{d_{\mbox{\footnotesize{sym}}}^{\mbox{\footnotesize{TIM}}}\times(1-t_l)}\}
=\frac{1}{1-t_l},
\end{align}
which is no larger than 2.

\section*{Appendix D: Proof of Theorem \ref{neighboring}} \label{neighboring_appendix}
First, consider the achievability. When $M$ is no larger than $S$, the achievable scheme is to treat all the medium interfering links as strong interfering links and apply the one-to-one alignment (see Theorem $6$ of \cite{Maleki_Cadambe_Jafar_index}). Note that this scheme falls into the category of TIM-TIN decomposition, where the TIN component contains no interfering links and the TIM component contains all the medium and strong interfering links. Otherwise, when $M$ is larger than $S$, we use the following decomposition to achieve the optimal symmetric GDoF value: let the TIN and TIM component contain all the medium interfering links and all the strong interfering links, respectively.  For the TIN component, the achievable symmetric GDoF value is $\frac{1}{2}$, and for the TIM component, the achievable symmetric GDoF value is $\frac{1}{S+1}$ \cite{Maleki_Cadambe_Jafar_index}. Therefore,  in the original network, the symmetric GDoF value $\frac{1}{2(S+1)}$ is achievable. 

Next, consider the converse. We start with a useful lemma.
\begin{lemma}
\label{3user_lemma}
Consider a 3-user interference channel within the QM-TIM(0,0.5) framework, where $i,j,k\in\{1,2,3\}$, $i\neq j$, $j\neq k$, and $i\neq k$. Denote by $l_{ij}$  the link between Transmitter $j$ and Receiver $i$, $\mathcal{M}$ the set of all medium interfering links, and $\mathcal{S}$ the set of all strong interfering links. If $l_{ij}\in \mathcal{S}$, and $l_{ki},l_{ik},l_{kj},l_{jk}\in \{\mathcal{S}\cup\mathcal{M}\}$, then the sum GDoF value of this channel is $1$.
\end{lemma}
\textit{Proof:} The achievability is straightforward. In the following we only consider the converse. Without loss of generality, we assume $i=1$, $j=2$, and $k=3$. To obtain the desired outer bound, we first set $\alpha_{21}=0$. This does not hurt the sum capacity because regardless of the channel strength level of the cross link $l_{21}$, we can always provide the message $W_1$ to Receiver $2$ through a genie and remove this interfering link. 

Without perfect channel knowledge at transmitters, the channel can be regarded as a compound channel, and its capacity is upper bounded by any possible channel realization. Recall that according to the definition of QM-TIM(0,0.5) given in Section \ref{sec_model}, for both strong and medium interfering links, their channel strength levels can be set as the threshold value 0.5. Consider a specific channel realization where $\alpha_{11}=\alpha_{22}=\alpha_{33}=\alpha_{12}=1$, $\alpha_{13}=\alpha_{31}=\alpha_{23}=\alpha_{32}=0.5$, and all the links have the same channel phase. The capacity of the original channel is upper bounded by this case. 

For any reliable decoding scheme, Receiver $1$ can always decode its own message $W_1$. After decoding $W_1$, Receiver $1$ can subtract it from its received signal and has the same signal as Receiver $2$. So Receiver 1 can also decode $W_2$. Now consider Transmitters $1$ and $2$. We find that they have the same channel vectors to Receiver $1$ and $3$. It implies that the sum capacity of the original channel is upper bounded by that of a 2-user interference channel with transmitters $\{T_{1,2}, T_{3}\}$ and receivers $\{R_1,R_3\}$, where $T_{1,2}$ is a combination of Transmitter 1 and 2. The sum-GDoF value of this 2-user interference channel (where both desired links have channel strength level 1 and both cross links have channel strength level 0.5) is known to be $1$ \cite{Tse_GDoF}. Therefore, we establish the desired outer bound.  \hfill $\blacksquare$ 

%\begin{figure}[h]
%\centering
%\subfigure[]{
%\includegraphics[width= 3.6 cm]{3user_lemma_1}
%\label{3user_lemma_1}}
%\subfigure[]{
%\includegraphics[width= 3.6 cm]{3user_lemma_2}
%\label{3user_lemma_2}}
%\subfigure[]{
%\includegraphics[width= 3.6 cm]{3user_lemma_3}
%\label{3user_lemma_3}}
%\caption[]{
%\subref{3user_lemma_1} The 3-user interference channel satisfying the condition in Lemma \ref{3user_lemma}, where the channel strength levels for the red solid line and blue dashed line are $1$ and $0.5$, respectively. \subref{3user_lemma_2} The 3-user interference channel after removing Receiver $2$. \subref{3user_lemma_3} In the final step, we combine Transmitter $1$ and $2$ into a new transmitter ($T_{12}$) and get a two-user interference channel, whose sum-GDoF value is equal to 1. }
%\label{3user_lemma_fig}
%\vspace{-0.2in}
%\end{figure}

%\bigskip
Now consider the following two cases.

\emph{Case I ($M\leq S$)}: For the converse, consider any consecutive $S+M+1$ users. Without loss of generality, assume that the user indices range from $1$ to $S+M+1$. For these users, we intend to prove the outer bound $d_1+d_2+...+d_{S+M+1}\leq 1$.

\begin{figure}[h]
\begin{center}
 \includegraphics[width=7 cm]{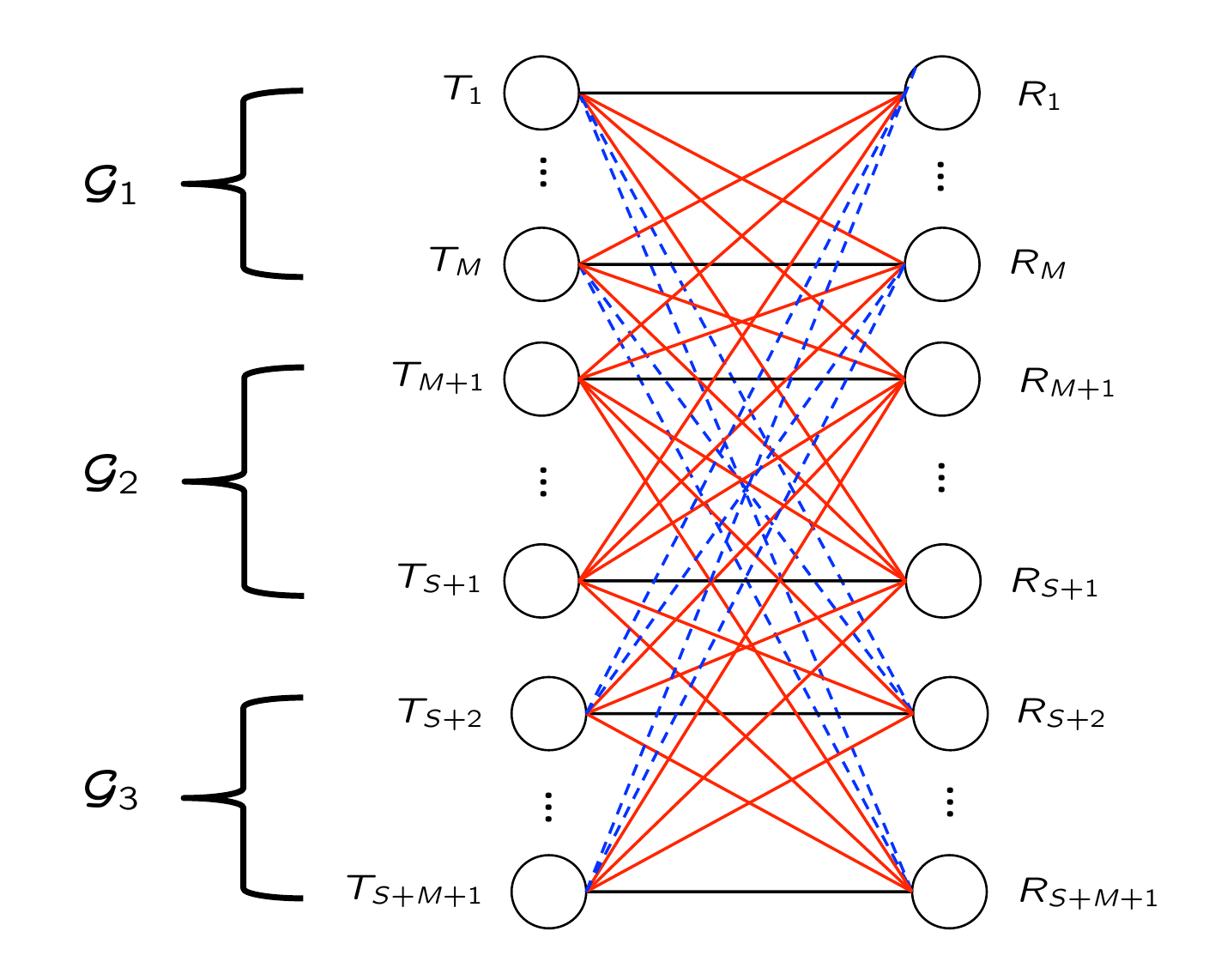}
 \caption{When $M\leq S$, for the converse we consider this $(S+M+1)$-user interference channel, where the channel strength levels of the red solid lines and blue dashed lines are $1$ and $0.5$, respectively.}
\label{neighbor_case1}
\end{center}
\vspace{-0.2in}
\end{figure}

Towards this end, first remove all the users other than the considered $S+M+1$ users, which does not hurt the sum capacity of users 1 to $S+M+1$. Next, in the remaining network, divide the $S+M+1$ users into three subgroups as shown in Fig.~\ref{neighbor_case1}:
    \begin{itemize}
    \item{$\mathcal{G}_1$}: this subgroup includes users $1$ to $M$;
    \item{$\mathcal{G}_2$}: this subgroup includes users $M+1$ to $S+1$;
    \item{$\mathcal{G}_3$}: this subgroup includes users $S+2$ to $S+M+1$.
    \end{itemize}

To derive the desired outer bound, consider the channel realization below. Assume that all the links have the same channel phase. For the direct links, recall that their channel strength levels are all equal to $1$. For the medium interfering links, we set their channel strength levels to be exactly $0.5$. For the strong interfering links, we set their channel strength levels to be either $1$ or $0.5$ as follows. For all the transmitters in $\mathcal{G}_1$, we assume that the cross links between them and the receivers in $\mathcal{G}_1$ and $\mathcal{G}_2$ are all with channel strength level $1$, while the cross links between them and the receivers in $\mathcal{G}_3$ are all with channel strength level $0.5$. Next, for the transmitters in $\mathcal{G}_2$, the cross links between them and all the receivers are with channel strength level $1$. Finally, for the transmitters in $\mathcal{G}_3$, the cross links between them and the receivers in $\mathcal{G}_2$ and $\mathcal{G}_3$ are all with channel strength level $1$, while the cross links between them and the receivers in $\mathcal{G}_1$ are all with channel strength level $0.5$.  

Now, note that in this network, all the receivers in the same subgroup $\mathcal{G}_i$ $i\in\{1,2,3\}$, are equipped with the same received signal. Thus removing all of them but one cannot hurt the sum capacity. Also note that for all the transmitters in the same subgroup $\mathcal{G}_i$, $i\in\{1,2,3\}$, they have the same channel vectors to all the remaining three receivers. Thus combining all the transmitters in each subgroup into one transmitter does not hurt the sum capacity either. Therefore, the network is reduced to a $3$-user interference channel where $\alpha_{13}=\alpha_{31}=0.5$ and all the other links are with channel strength level value 1. According to Lemma \ref{3user_lemma}, the sum-GDoF value of this network is $1$, which establishes the desired outer bound.

%\begin{figure}[h]
%\begin{center}
 %\includegraphics[width= 6 cm]{neighbor_case1_final}
 %\caption{When $M\leq S$, after removing redundant receivers and combining equivalent transmitters, we get a 3-user interference channel, where $T_{i:j}$ denotes the transmitter obtained by combining the transmitters with indices from $i$ to $j$.}
%\label{neighbor_case1_final}
%\end{center}
%\vspace{-0.1in}
%\end{figure}

\begin{figure}[h]
\begin{center}
 \includegraphics[width= 7 cm]{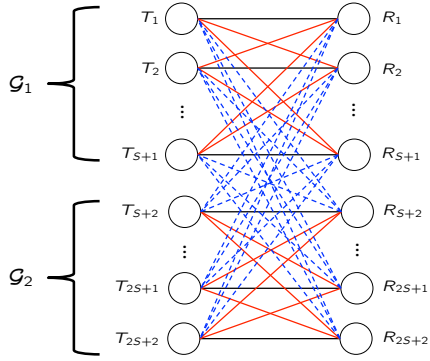}
 \caption{When $M> S$, for the converse we consider this $2(S+1)$-user interference channel, where the channel strength levels of the red solid lines and blue dashed lines are $1$ and $0.5$, respectively.}
\label{neighbor_case2}
\end{center}
\vspace{-0.2in}
\end{figure}

\emph{Case II ($M> S$)}: For the converse, consider any consecutive $2(S+1)$ users. Without loss of generality, assume the user indices range from $1$ to $2(S+1)$. For these users, we intend to show $d_1+d_2+...+d_{2(S+1)}\leq 1$. Similar to the previous case, we first remove all the other users.  In the remaining network, divide the $2(S+1)$ users into two subgroups as shown in Fig.~\ref{neighbor_case2}:
    \begin{itemize}
    \item{$\mathcal{G}_1$}: this subgroup includes users $1$ to $S+1$;
    \item{$\mathcal{G}_2$}: this subgroup includes users $S+2$ to $2(S+1)$.
    \end{itemize}

 Again, assume that all the links have the same channel phase. For the direct links, their channel strength levels are all $1$. For the medium interfering links, we set their channel strength levels to be exactly $0.5$. Next, we set the channel strength levels of the strong interfering links to be either $1$ or $0.5$ as follows. For transmitters in each subgroup $\mathcal{G}_i$, the cross links between them and the receivers in the same subgroup $\mathcal{G}_i$ are all with channel strength level $1$, and the cross links between them and all the receivers in the other subgroup $\mathcal{G}_j$ are with channel strength level $0.5$, where $i,j\in\{1,2\}$ and $i\neq j$. Removing all the receivers but one in each subgroup $\mathcal{G}_i$, $i\in\{1,2\}$, cannot hurt the sum capacity. Combining all the transmitters in each subgroup $\mathcal{G}_i$, $i\in\{1,2\}$, into one transmitter cannot hurt the sum capacity either. Finally, we end up with a $2$-user interference channel with $\alpha_{11}=\alpha_{22}=1$, $\alpha_{12}=\alpha_{21}=0.5$ and sum-GDoF value 1 \cite{Tse_GDoF}, which leads to the desired outer bound.

%\begin{figure}[h]
%\begin{center}
 %\includegraphics[width= 6 cm]{neighbor_case2_final}
 %\caption{When $M> S$, after removing redundant receivers and combining equivalent transmitters, we get a $2$-user interference channel.}
%\label{neighbor_case2_final}
%\end{center}
%\vspace{-0.2in}
%\end{figure}

  \section*{ Appendix E: Proof of Theorem \ref{T2_convergence}}\label{app_T2}
 As the sum-GDoF of an interference channel must be upper bounded by a finite value, to prove this theorem, we only need to show that the achievable sum-GDoF via the ZEST algorithm monotonically increases after each update, i.e., $\overrightarrow{d}_{\Sigma}^{(m)}\leq \overleftarrow{d}_{\Sigma,\mbox{\small{switch}}}^{(m)}\leq  \overleftarrow{d}_{\Sigma}^{(m)}\leq \overrightarrow{d}_{\Sigma,\mbox{\small{switch}}}^{(m)} \leq \overrightarrow{d}_{\Sigma}^{(m+1)}$. Towards this end, we show that the GDoF tuple obtained in each step satisfies 
  \begin{align} \label{key_T2}
  \overrightarrow{\bf{d}}^{(m)}\overset{(a)}\leq \overleftarrow{\bf{d}}_{\mbox{\small{switch}}}^{(m)}\overset{(b)}\leq  \overleftarrow{\bf{d}}^{(m)}\overset{(c)}\leq \overrightarrow{\bf{d}}_{\mbox{\small{switch}}}^{(m)} \overset{(d)}\leq \overrightarrow{\bf{d}}^{(m+1)},
  \end{align}
  where (b) and (d) follow from Lemma \ref{lemma_ZF} directly, as in these two steps, the receiver is updated to the ZF-SC receiver that achieves the maximal GDoF.

Next, consider (a). Let $B=\sum_{k=1}^Kb_k$.  In the $m$-th iteration, define an indicator function
  \begin{align}
  \mbox{I}_{\{{|\overrightarrow{\mathbf{u}}_{k,l}^{(m)\dag}\overrightarrow{\mathbf{v}}_{j,s}^{(m)}|}\neq 0\}}=\left\{\begin{array}{cc}
           1, & |\overrightarrow{\mathbf{u}}_{k,l}^{(m)\dag}\overrightarrow{\mathbf{v}}_{j,s}^{(m)}|\neq 0 \\
           0, & |\overrightarrow{\mathbf{u}}_{k,l}^{(m)\dag}\overrightarrow{\mathbf{v}}_{j,s}^{(m)}|= 0
          \end{array}
          \right.
  \end{align}
Next, define $ G_{k,l}^{j,s}=\alpha_{kj}\mbox{I}_{\{{|\overrightarrow{\mathbf{u}}_{k,l}^{(m)\dag}\overrightarrow{\mathbf{v}}_{j,s}^{(m)}|}\neq 0\}}$, which is the effective channel strength level from data stream $s$ of User $j$ to data stream $l$ of User $k$ in the original channel.  Also define a $B\times B$ matrix $\mathbf{G}\bigg(\sum_{n=1}^{k-1}b_n+l,\sum_{m=1}^{j-1}b_m+s  \bigg) = G_{k,l}^{j,s}$.

Recall that a successive cancellation procedure is adopted at each receiver. According to the ZEST algorithm given in Section \ref{sec_dis},  without loss of generality, we have assumed that the cancellation is taken in the lexicographic order. Therefore, for Receiver $k\in[K]$, the effective channel strength level from data stream $p$ of User $k$ to data stream $q$ of User $k$ is $0$, for $p,q\in[b_k]$ and $p< q$. Set the corresponding entries of $\mathbf{G}$ as $0$, i.e.,
\begin{align}
\mathbf{G}\bigg(\sum_{n=1}^{k-1}b_n+q,\sum_{n=1}^{k-1}b_n+p\bigg)=0,~~\forall k\in[K],~\forall p,~q\in[b_k],~ p<q,
\end{align}
and denote the obtained matrix by $\overrightarrow{\mathbf{G}}$. Next, for the $K$-user original channel in the $m$-th iteration with beamforming vectors $\overrightarrow{\mathbf{v}}_{j,s}^{(m)}$ and ZF-SC receiving vectors $\overrightarrow{\mathbf{u}}_{k,l}^{(m)}$, we construct a counterpart $B$-user interference channel with the channel strength level matrix $\overrightarrow{\mathbf{G}}$, which is denoted by $\mathcal{IC}_o$. For $\mathcal{IC}_o$, $\overrightarrow{\mathbf{G}}(j,i)$ denotes the channel strength level from Transmitter $i$ to Receiver $j$. Assume that in $\mathcal{IC}_o$, the allocated power to Transmitter $i$  is $\overrightarrow{r}_i=\overrightarrow{r}_{j,s}^{(m)}$ where $i=\sum_{l=1}^{j-1}b_l+s$.  By treating interference as noise at each receiver, we obtain the achievable GDoF tuple of  $\mathcal{IC}_o$ $(d_{1,o},...,d_{B,o})$ and $\sum_{i=i_j}^{i_j'} d_{i,o}=n\times\overrightarrow{d}_{j}^{(m)}$, where $i_j=\sum_{l=1}^{j-1}b_l+1$, $i_j'=\sum_{l=1}^{j}b_l$, and $\overrightarrow{d}_{j}^{(m)}$ is the $j$-th entry of $\overrightarrow{\bf{d}}^{(m)}$.

Similarly, for the reciprocal channel in the $m$-th iteration with beamforming vectors $\overleftarrow{\mathbf{v}}_{j,s}^{(m)}=\overrightarrow{\mathbf{u}}_{j,s}^{(m)}$ and receiving vectors $\overleftarrow{\mathbf{u}}_{k,l}^{(m)}=\overrightarrow{\mathbf{v}}_{k,l}^{(m)}$, we construct a counterpart $B$-user interference channel with the channel strength level matrix $\overleftarrow{\mathbf{G}}=\overrightarrow{\mathbf{G}}^T$, which is the reciprocal channel of $\mathcal{IC}_o$ and denoted by $\mathcal{IC}_{r}$.\footnote{Note that the new channel $\mathcal{IC}_r$ with the channel strength level matrix $\overrightarrow{\mathbf{G}}^T$ corresponds to the reciprocal channel in the $m$-th iteration where the successive cancellation for each user is taken in the \emph{reverse} lexicographic order.} Assume that in $\mathcal{IC}_r$,  the allocated power to Transmitter $i$ is $\overleftarrow{r}_i=\overleftarrow{r}_{j,s}^{(m)}$ where $i=\sum_{l=1}^{j-1}b_l+s$.  By treating interference as noise at each receiver, we obtain the achievable GDoF tuple of $\mathcal{IC}_r$ $(d_{1,r},...,d_{B,r})$ and $\sum_{i=i_j}^{i_j'}d_{i,r}=n\times\overleftarrow{d}_{j,\mbox{\small{switch}}}^{(m)}$, 
where $\overleftarrow{d}_{j,\mbox{\small{switch}}}^{(m)}$ is the $j$-th entry of $\overleftarrow{\bf{d}}_{\mbox{\small{switch}}}^{(m)}$. According to Lemma \ref{T1_dual}, we have
\begin{align}
\sum_{i=i_j}^{i_j'} d_{i,o}\leq  \sum_{i=i_j}^{i_j'}d_{i,r} \Rightarrow \overrightarrow{d}_{j}^{(m)} \leq\overleftarrow{d}_{j,\mbox{\small{switch}}}^{(m)},~~\forall j\in[K],
\end{align}
and hence prove (a). The proof of (c) follows similarly. Therefore, we establish (\ref{key_T2}) and complete the proof.

\end{document}